\documentclass[aip,amsmath,amssymb, reprint,numerical]{revtex4-1}
\usepackage[version=4]{mhchem} % Formula subscripts using \ce{}
\usepackage{booktabs}
\usepackage{braket}
\usepackage[T1]{fontenc}
\usepackage{mathptmx,xcolor}
\usepackage[utf8]{inputenc}
\usepackage{cleveref}
\usepackage{multirow}
\usepackage{rotating}
\usepackage{graphicx}
\usepackage{siunitx}
\usepackage{subcaption}
\usepackage{amsfonts,amsmath}

\newcommand*{\mc}{\multicolumn}

\usepackage{array}
\newcolumntype{H}{>{\setbox0=\hbox\bgroup}c<{\egroup}@{}}

\AtBeginDocument{
\heavyrulewidth=.08em
\lightrulewidth=.05em
\cmidrulewidth=.03em
\belowrulesep=.65ex
\belowbottomsep=0pt
\aboverulesep=.4ex
\abovetopsep=0pt
\cmidrulesep=\doublerulesep
\cmidrulekern=.5em
\defaultaddspace=.5em
}
\graphicspath{{Images/}}

\begin{document}

\title{Mixed ramp-Gaussian basis sets for core-dependent properties: STO-RG and STO-R2G for Li-Ne.}

\author{Claudia S. Cox}

% List of institutions
\affiliation{School of Chemistry, University of New South Wales, Kensington, Sydney, 2052 Australia}

\author{Juan Camilo Zapata}

\affiliation{Departamento de Ciencias Qu\'imicas, Universidad Icesi, Cali, Valle del Cauca, Colombia}

\author{Laura K. McKemmish$^{1}$} \email{l.mckemmish@unsw.edu.au}

\affiliation{School of Chemistry, University of New South Wales, Kensington, Sydney, 2052 Australia}


\begin{abstract}
%\alert{Claude can you rework these two sentences to introduce need for ramps in core electron quantum chemistry: 
%Computational quantum chemistry relies on the ability of basis sets to describe the electron wavefunction in the area of interest. Describing core electronic properties is challenging as Gaussian basis sets lack an electron-nuclear cusp. }

\textbf{Peer-reviewed publication \url{https://doi.org/10.1071/CH19466}}

The traditional Gaussian basis sets  used in modern quantum chemistry lack an electron-nuclear cusp, and hence struggle to accurately describe core electron properties. A recently introduced novel type of basis set, mixed ramp-Gaussians, introduce a new primitive function called a ramp function which addresses this issue. 

This paper introduces three new mixed ramp-Gaussian basis sets - STO-R, STO-RG and STO-R2G, made from a linear combination of ramp and Gaussian primitive functions - which are derived from the single-core-zeta Slater basis sets for the elements Li to Ne. This derivation is done in an analogous fashion to the famous STO-$n$G basis sets. The STO-RG basis functions are found to outperform the STO-3G basis functions and STO-R2G outperforms STO-6G, both in terms of wavefunction fit and other key quantities such as the one-electron energy and the electron-nuclear cusp. 

The second part of this paper performs preparatory investigations into how standard all-Gaussian basis sets can be converted to ramp-Gaussian basis sets through modifying the core basis functions. Using a test case of the 6-31G basis set for carbon, we determined that the second Gaussian primitive is less important when fitting a ramp-Gaussian core basis function directly to an all-Gaussian core basis function than when fitting to a Slater basis function. Further, we identified the basis sets that are single-core-zeta and thus should be most straightforward to convert to mixed ramp-Gaussian basis sets in the future. % are the first targets of future  be suitable for straightforward  with a vie we analysed a large set of all-Gaussian basis sets  %While a ramp and two Gaussians were required to model the core region of a Slater accurately, only one ramp and one Gaussian were needed to model 6-31G well. With an eye of creating more derived mixed ramp-Gaussian basis sets, an analysis was conducted on traditional all-Gaussian basis sets. Unlike these basis sets, mixed ramp-Gaussian basis sets have the potential to revolutionise the computational prediction of core electronic properties.
\vspace{4em}

\end{abstract}
\maketitle

%\alert{CHECK consistency of L2 being multipled by 1000 not 10,000!!}
\renewcommand{\arraystretch}{1.2}

% \begin{figure}[h]
%     \includegraphics[width=0.5\textwidth]{Cusp.png}
%     \caption{Wav}
%     \label{fig:cusp}
% \end{figure}

% \begin{figure}
% %  \begin{subfigure}[b]{0.4\textwidth}
%     \includegraphics[width=\textwidth]{diffwavemetrics$Z_\textrm{eff}$10.png}
% %  \end{subfigure}
% %  \begin{subfigure}[b]{0.4\textwidth}
%     \includegraphics[width=\textwidth]{diffwavemetrics$Z_\textrm{eff}$40.png}
% %  \end{subfigure}
% %    \begin{subfigure}[b]{0.4\textwidth}
% \end{figure}
% \begin{figure}
%     \includegraphics[width=\textwidth]{diffprobmetrics$Z_\textrm{eff}$10.png}
% %  \end{subfigure}
% %    \begin{subfigure}[b]{0.4\textwidth}
%     \includegraphics[width=\textwidth]{diffprobmetrics$Z_\textrm{eff}$40.png}
% %  \end{subfigure}
%   \end{figure}

%\end{document}

\section{Introduction}
Over the last half century, computational chemistry has grown from a fringe research area to an important component of almost all chemical research. Chemists can now routinely model structure \cite{16BrSaSu}, thermochemistry \cite{thermochem,12PeFeDi} and reactivity \cite{reactivity} of chemical compounds. As acknowledged by Pople's 1998 Nobel Prize \cite{pople_lecture}, this success can be largely attributed to the computationally time efficient way in which Gaussian basis functions model valence electrons,  alongside density functional theory \cite{DFT_zoo} as a computationally inexpensive but accurate method of approximating the Schr\"odinger equation. 

Even though chemical structure and reactivity is mainly dependent on valence electrons \cite{valence_electrons}, many important spectroscopic properties are dependent on core electrons.  These properties include nuclear magnetic resonance (NMR)\cite{nmr_review} chemical shifts \cite{pcS,pcSseg-n} and spin-spin couplings \cite{pcJ}, electron paramagentic resonance (EPR) spectroscopy and hyperfine couplings \cite{epr_review}, Auger \cite{auger_review} and M\"ossbauer spectroscopy \cite{gibb_mossbook} and all interactions of molecules with X-rays \cite{xray_review1, xray_review2,pcX}. 

The quality of a computational chemistry description of the valence and core electron region depends on the choice of basis set. Mathematically, molecular orbitals, $\phi$, are described as a linear combination of basis functions, $\chi$, i.e.
\begin{equation}
    \phi = \sum_{i}^{M}k_{i}\chi_{i}
    \label{basisset}
\end{equation}
 where the coefficients $k_{i}$ represent the contribution of each basis function in the molecular orbital representation and are determined by solving the Hartree-Fock or Density Functional Theory (DFT) equations. The selection of these basis functions is crucial, as their behaviour must efficiently and effectively represent the electron distribution and hence accurately determine the investigated molecular properties \cite{jensen_book}. Generally, basis functions, $\chi_i$, can be comprised of one or more primitive basis functions, e.g. the contracted core basis function in the 6-31G basis set contains six primitive Gaussian basis functions. 
 
 %Basis sets generally describe the valence region well but are There are a few types of basis functions, Slater functions, Gaussian functions and ramp functions (with relevant notation given in \Cref{tab:notation}) that are used for molecular systems. 
 
 \begin{table}
\centering

     \caption{Three key types of basis functions, and notation used in this work. }
     \label{tab:notation}
\begin{tabular}{lcccccccc}
     \toprule
         Type & Radial Var. & \mc{2}{c}{General $\ell,m$}  & \mc{2}{c}{$\ell=0$}  \\
        \cmidrule(r){1-1} \cmidrule(r){2-2} \cmidrule(r){3-4} \cmidrule(r){5-6}
%         & & Gen. & Spec. & Generic & Specific \\
  %       \midrule
         Slater & $Z_\textrm{eff}$ & $\mathcal{S}$ & $\mathcal{S}_{Z_\textrm{eff}\ell m}$ & $\mathbb{S}$ &  $\mathbb{S}_{Z_\textrm{eff}}$ \\ 
         Gaussian & $\alpha$  & $\mathcal{G}$ & $\mathcal{G}_{\alpha \ell m}$ &$s$ & $s_\alpha$ \\
         Ramp & $n$ & $\mathcal{R}$ & $\mathcal{R}_{n\ell m}$ & $S$  & $S_n$ \\
         \bottomrule
     \end{tabular}
\end{table}

 This paper considers three types of basis functions in depth - Slaters, Gaussians and ramps. Relevant notation for each of these basis function types is given in \Cref{tab:notation}, with different notation for the generic function ($\mathcal{S},\mathcal{G},\mathcal{R}$) and the s-radial component ($\mathbb{S},s,S)$. Though in principle a contracted core basis function could be a linear combination of any or all of these three types of basis functions, in practice we only consider contracted basis functions with a single type of basis function or with a single ramp and Gaussians, e.g. $c_1 S_n + c_2 s_{\alpha_1}  + c_3 s_{\alpha_2}$.

%\alert{May want to discuss numerical atomic orbital basis functions - see https://arxiv.org/pdf/1805.12225.pdf}

The earliest basis sets utilised hydrogenic orbitals as basis functions; these are known as Slater functions \cite{slater1930atomic}  (denoted herein by $\mathcal{S}$). The Slater function with angular momentum quantum numbers $\ell$ and $m$ and exponent $\alpha$ has the mathematical form
\begin{equation}
    \mathcal{S}_{\alpha \ell m} (\textbf{r}) = N^{\mathcal{S}}_{\alpha \ell m} Y_{\ell m}(\theta, \phi) r^{\ell} e^{-\alpha r}
    \label{slater}
\end{equation}
with a normalisation factor $N^{\mathcal{S}}_{\alpha \ell m}$ given by
\begin{equation}
    N^{\mathcal{S}}_{\alpha \ell m} = (2\alpha)^{\ell} \sqrt{\frac{2\alpha}{(2\ell)!}}
    \label{nomrS}
\end{equation}
where $Y_{\ell m}(\theta, \phi)$ is a spherical harmonic function.
%Unlike Gaussian functions, Slater functions do have the correct electron-nuclear cusp behaviour.
Slater functions are very efficient at describing electron density, meaning that few basis functions are needed for accurate description of molecular systems. However, almost 70 years of research (e.g. \cite{sto_comp_review, sto_comp1, sto_comp2, sto_comp3, sto_comp4}) has failed to find efficient methods of calculating the two-electron multi-centre integrals that arise when using Slater functions for quantum chemistry computations \cite{sto_inefficient}. These integral difficulties mean that today Slater functions are restricted to non-mainstream programs such as Amsterdam Density Functional (ADF) \cite{adf}, though notably these programs are still used for applications like predicting NMR spectral parameters.

\begin{figure}
\centering
    \includegraphics[width=0.3\textwidth]{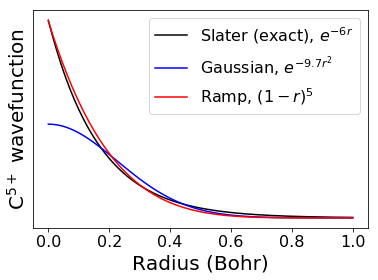}
    \caption{Single Slater, Gaussian and ramp wavefunctions for modelling the \ce{C^{5+}} wavefunction. }
    \label{fig:cusp}
\end{figure}

In 1950, Boys \cite{Boys50} proposed the use of Gaussian functions as atomic basis functions for quantum chemistry and the foundation of modern quantum chemistry was established. Mathematically, the Gaussian basis function, denoted herein by $\mathcal{G}$, with exponent $\alpha$ and angular momentum quantum numbers $\ell$ and $m$ is given by
\begin{equation}
    \mathcal{G}_{\alpha \ell m} (\textbf{r}) = N^{\mathcal{G}}_{\alpha \ell m} Y_{\ell m}(\theta, \phi) r^{\ell} e^{-\alpha r^{2}},
    \label{Gaussian}
\end{equation}
where ${N}^{\mathcal{G}}_{\alpha \ell m}$, the normalisation factor, has the mathematical form
\begin{equation}
    N^{\mathcal{G}}_{\alpha \ell m} = \sqrt{\frac{(\ell+1)!(8\alpha)^{\ell+3/2}}{(2\ell+2)!\sqrt{\pi}}}
    \label{normGau}
\end{equation}
These Gaussian functions are sub-optimal for modelling electron wavefunctions as they (1) lack an electron-nuclear cusp, i.e. for Gaussian functions $\frac{d\chi}{dr}\big|_{r=0} = 0$, as shown visually in \Cref{fig:cusp} and (2) decay too rapidly at large $r$. These deficiencies means that per basis function Gaussian functions have inferior performance than to Slater functions \cite{gaussians}. Nevertheless, the ease of computing two-electron integrals \cite{gill1994simple} means that Gaussian basis functions are overwhelming utilised throughout modern quantum chemistry for molecular systems.

The first popular Gaussian-type basis sets were the STO-$n$G basis sets \cite{sto-nG}, developed by fitting a Slater function as a sum of $n$ Gaussians-type functions by maximises the overlap (i.e. minimising $1-\Braket{\mathcal{S}|\mathcal{G}}$) between both functions. Gaussian basis functions  proved so useful \cite{gaussians} that subsequently a large number of all-Gaussian basis sets were developed, from the famous Pople style basis sets such, as STO-$n$G and 6-31G \cite{3-21G, 6-31G, 6-31Gstar, 6-311G_star, 6-31Gd}, to more modern families like the Dunning \cite{cc-pVnZ_B-Ne,cc-pCVnZ,cc-pwCVTZ} and Jensen \cite{pc-n, pcseg-n, pcSseg-n, pcJ, pcX, pcS} basis sets. The range of atomic-centred Gaussian basis sets have been reviewed extensively \cite{atomic_orbitals,Nagy2017BasisChemistry} with most popular basis sets available online at Basis Set Exchange \cite{BSE1,BSE2}.%,BSE3}.

Recently \cite{12McGi}, it was shown that the inability of Gaussian basis functions to model nuclear-electron cusps was more significant in reducing their accuracy than their incorrect long-range decay. This effect can be seen by examining the breakdown of number of primitive Gaussian functions used in typical all-Gaussian basis sets; even those designed to model valence chemistry (such as the extremely popular Pople basis set 6-31G* \cite{6-31Gstar}) use at least six primitive Gaussian basis functions in the core region (with very high exponents).

When describing core-dependent properties computationally, all-Gaussian basis sets  with a large number of partially or fully uncontracted $s$-primitives basis functions have been developed \cite{wachters_basis, cp_ppp, pcS, pcJ,pcX}.  Nevertheless, currently these basis sets must still rely on Gaussian-type functions or forfeit computational efficiency.  

Alternatives to specialised all-Gaussian basis sets include numerical atomic orbitals (NAOs) with flexible shape and correct near core behaviour \cite{num_ato}. Calculations have been performed for molecular systems with more than 1000 atoms \cite{num_ato} with the general purpose NAO-VCC-$n$Z basis sets \cite{13ZhReRi.NAO} and the specialised   NAO-J-$n$ basis sets \cite{num_ato}. NMR chemical shifts calculated using the general-purpose NAO-VCC-$n$Z basis set converged much more quickly to the basis set limit than general-purpose all-Gaussian basis sets, at approximately the same rate as the specialised Jensen pcS-$n$ basis set. The NAO-J-$n$ basis set has similar  convergence properties to the analogous specialised Jensen pcJ-$n$ basis sets \cite{num_ato}. 

In 2014, McKemmish \emph{et. al.} \cite{14McGiGi} proposed a fundamentally new type of basis set,  named mixed ramp-Gaussian basis sets, that have the potential to revolutionise our ability to computationally model core-electrons. These basis sets retain Gaussians to describe the valence electron distribution, but use a new type of basis function, the ramp (denoted by $\mathcal{R}$) \cite{ramps_1, ramps_2}, which has an electron-nuclear cusp, to describe the core-electron distribution. 
\textcolor{black}{ A ramp function with degree $n$ and angular momentum quantum numbers $\ell$ and $m$ is defined by 
 \begin{equation}
    \mathcal{R}_{n\ell m} (\textbf{r}) =  \left\{ \begin{array}{lcc}
             N^{\mathcal{R}}_{n \ell m} Y_{\ell m}(\theta, \phi) r^{\ell}(1-r)^{n} & : & r \leq 1\\
             \\ 0 & : & r > 1
             \end{array}
   \right.
    \label{ramp}
\end{equation}
where $N^{\mathcal{R}}_{n\ell m}$ is the normalisation factor given by
\begin{equation}
    N^{\mathcal{R}}_{n \ell m} = \sqrt{\frac{(2n+2\ell+3)!}{(2n)!(2\ell+2)!}}
    \label{normR}
\end{equation}
where $n$ is the degree of the ramp, not the principal quantum number. The support of the function is chosen as 1 Bohr because this enables two-centre ramp-ramp shell pairs to be avoided in most practical calculations (as bond lengths between non-hydrogen atoms are typically greater than 2 Bohr). This choice dramatically decreases the complexity of the two-electron integrals and enables calculation times with ramp-Gaussian basis sets to be competitive  with all-Gaussian basis sets \cite{15Mc}. In this paper, the degree of the ramp, $n$, is constrained to be an integer, as this makes two-electron integrals easier and avoids unbound derivatives at $r = 1$ Bohr.}

As shown in \Cref{fig:cusp}, the ramp function has a electron-nuclear cusp, as it has a discontinuous first derivative at $r=0$, allowing these functions to more accurately describe the behaviour of electrons at and near the nucleus.

 The mixed ramp-Gaussian basis set R-31G had been developed by fitting the six-fold contracted core basis functions ($\mathcal{G}$) in 6-31G, with two-fold contracted core basis functions ($\mathcal{RG}$) in R-31G, one ramp and one Gaussian by minimising $\braket{\mathcal{RG}^2-\mathcal{G}^2|\mathcal{RG}^2-\mathcal{G}^2}$. The R-31G basis set has been demonstrated to deliver cc-pVQZ quality results for the electron density at the nuclus despite its small size \cite{15McGi}. Furthermore, an initial implementation has shown that integrals for mixed ramp-Gaussian basis sets can be evaluated in comparable times to all-Gaussian basis sets, using a series of effective evaluation techniques described in \cite{15Mc}.

To realise the benefits of these new types of basis sets, we need to specify the parameters of a variety of mixed ramp-Gaussian basis sets of different sizes and to model different properties. 
A full optimisation of these basis set parameters will be time consuming, so it is useful to start by "rampification" of pre-existing basis sets, ie. replacing (or rampifying) the existing core basis functions of the parent basis set with a mixed ramp-Gaussian basis function whilst retaining existing all-Gaussian valence basis functions. 

In \Cref{sec:Slater}, we develop the STO-R$n$G basis sets for Li-Ne by rampifying a STO core basis function whilst adopting the STO-$(n+4)$G valence basis functions. We consider the best metric to use when rampifying basis functions, and the influence of $n$ on the basis set quality.

In \Cref{sec:Gauss}, we turn our attention to preparing for the rampification of  all-Gaussian basis sets. The pre-existing R-31G basis set is compared against the novel R2-31G, STO-RG and STO-R2G basis set to explore the generality of the mixed ramp-Gaussian core basis function parameters. Further, we categorise common general-purpose and specialised all-Gaussian basis sets as single-core-zeta or multiple-core-zeta based on the relative contribution of the $s$ basis functions to the $1s$ orbital in the neutral carbon atom and the similarity of the basis functions to $\mathbb{S}_{5.67}$.

%\alert{Explain the terms "rampification" and "rampify" in the introduction somewhere}

\begin{table*}
    \sisetup{round-mode=places,retain-explicit-plus}
    \caption{Fit parameters and fit quality for $c_1 S_n + c_2 s_\alpha$ basis function fit to $\mathbb{S}_{5.67}$ and $\mathbb{S}_{16.43}$ using metrics 1, 2 and 3 (see text for definitions).  $\mathcal{L}_1$, $\mathcal{L}_2$ and $\mathcal{L}_3$ are the residuals calculated using each metric between the fitted ramp-Gaussian function and the original Slater function. $\Delta \chi(0)$ is the value of the fitted ramp-Gaussian function at the nucleus minus the value of the original Slater function at the nucleus. $\frac{\chi^\prime(r)}{\chi(r)}\Big|_{r=0}$ is the cusp of the fitted ramp-Gaussian function minus the cusp of the original Slater function. $\Delta \mathbb{E}(\chi_1)$ is the energy of the hydrogenic atom with nuclear charge $Z_\textrm{eff}$ as calculated by the fitted ramp-Gaussian function minus the true energy, and is given by  $\Delta \mathbb{E}(\chi_1) = \Braket{\chi_1|-\frac{Z_{\textrm{eff}}}{r} - \frac{\nabla^2}{2}|\chi_1} + \frac{Z_\textrm{eff}^2}{2}$. }
    \label{tab:metrics}
    \centering
    \begin{tabular}{lS[round-precision=7,table-parse-only]S[round-precision=3]S[round-precision=7,table-parse-only]S[round-precision=3]S[round-precision=3]S[round-precision=3]S[round-precision=3]S[round-precision=3]S[round-precision=3]S[round-precision=3]S[round-precision=3]S[round-precision=3]S[round-precision=3]S[round-precision=3]}
    \toprule
        & \mc{4}{c}{Fit Parameters} & \mc{6}{c}{Fit Quality} \\
        \cmidrule(r){2-5} \cmidrule(r){6-11}
       \textbf{Fitting metric}  & \mc{1}{c}{$c_1$} & \mc{1}{c}{$n$} & \mc{1}{c}{$c_2$} & \mc{1}{c}{$\alpha$} &  \mc{1}{c}{$1000\mathcal{L}_1$} & \mc{1}{c}{$1000\mathcal{L}_2$} & \mc{1}{c}{$1000\mathcal{L}_3$} &   \mc{1}{c}{$\Delta \chi(0)$} & \mc{1}{c}{$ \Delta \frac{\chi^\prime(r)}{\chi(r)}\Big|_{r=0}$} & \mc{1}{c}{$\Delta \mathbb{E}(\chi_1)$}  \\
        \midrule
    \mc{3}{l}{Fit to Carbon $1s$ basis function, $\mathbb{S}_{5.67}$} \\
               \vspace{-0.8em} \\
        \textbf{metric 1}, $\mathcal{L}_1$ & 0.493882726 & 7 & 0.2974791061 & 4.896756641  & 0.3750236711 & 0.8866115196 & 21.35378025 & -0.04935229473 & 0.07549776824 & 0.03160317054\\
        \textbf{metric 2}, $\mathcal{L}_2$ & 0.5327114207 & 7 & 0.2847065094 & 4.283686466  & 10.69752187 & 0.3738261845 & 30.37365293 & -0.359566592 & 0.2861992186 & 0.03337475729 \\
        \textbf{metric 3}, $\mathcal{L}_3$ & 0.5091346993 & 7 & 0.2916348459 & 4.725014834  & 2.436659597 & 0.6644612306 & 17.47396221 & -0.1810985292 & 0.1528665685 & 0.02563330096 \\
        \vspace{-0.5em} \\
        \mc{3}{l}{Fit to Chlorine $1s$ basis function, $\mathbb{S}_{16.43}$} \\
        \vspace{-0.8em} \\
        \textbf{metric 1}, $\mathcal{L}_1$ & 0.7473718777 & 18 & 0.1485102656 & 32.8532134  & 3.79557169 & 0.3992167723 & 13.57756078 & -0.1953049238 & 0.2072403144 & 0.1146758064 \\
        \textbf{metric 2}, $\mathcal{L}_2$ & 0.7791493614 & 18 & 0.1375546398 & 26.57296989  & 99.55402822 & 0.08972987568 & 16.85962765 & -1.079073398 & 0.5180634353 & 0.07490992524 \\
        \textbf{metric 3}, $\mathcal{L}_3$ & 0.7513549273 & 18 & 0.1465900246 & 32.54892427  & 6.566536681 & 0.3821333378 & 12.33742879 & -0.3247494084 & 0.238778259 & 0.1070477424\\
        \bottomrule
    \end{tabular}
\end{table*}

\begin{figure*}
    \centering
    \begin{subfigure}[t]{0.45\textwidth}
        \centering
        \includegraphics[width=\textwidth]{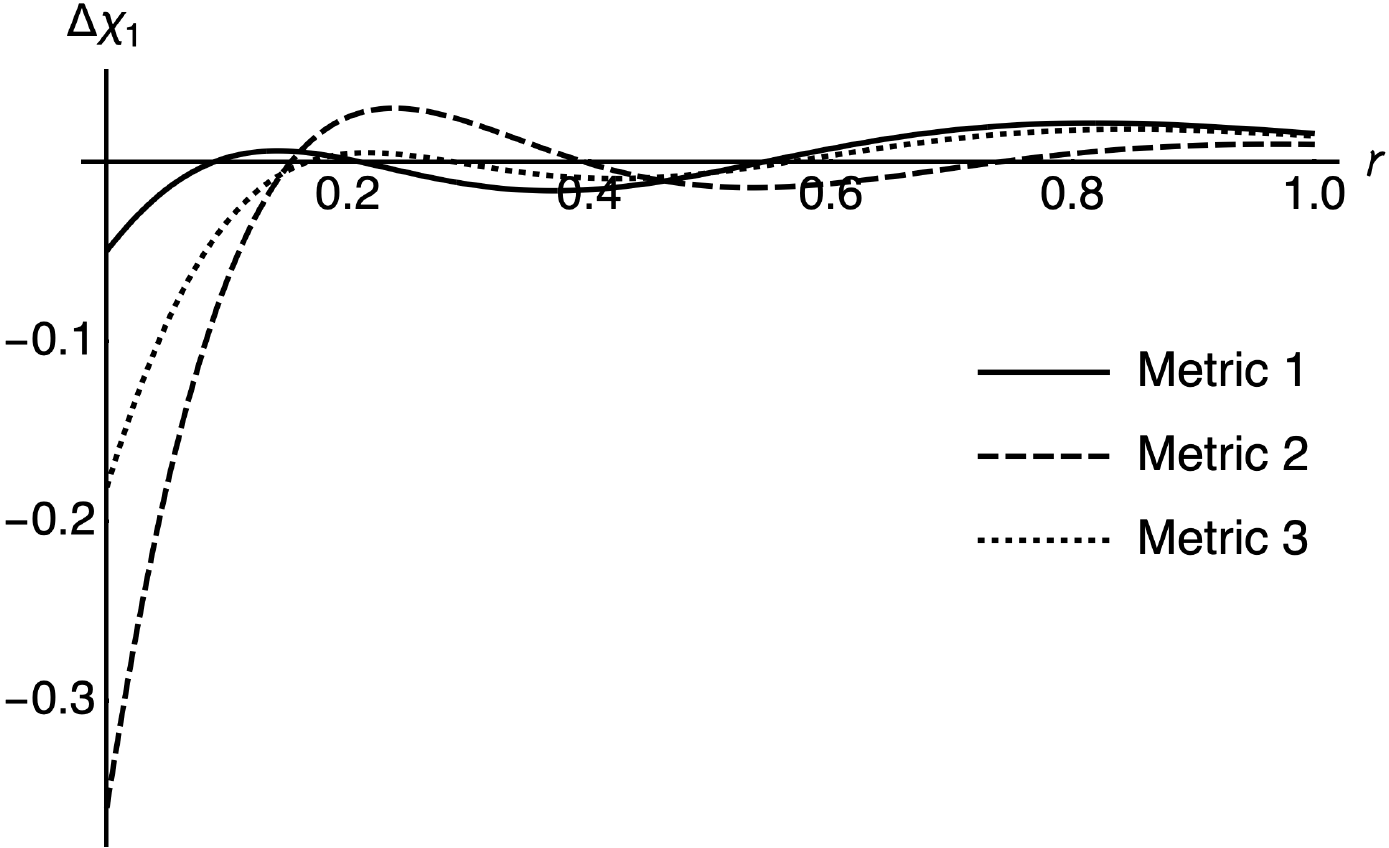}
        \caption{(a) $\Delta \chi_1$, $Z_\textrm{eff} = 5.67$, i.e. Carbon}
    \end{subfigure}%
    \hspace{1em}
    \begin{subfigure}[t]{0.45\textwidth}
        \centering
        \includegraphics[width=\textwidth]{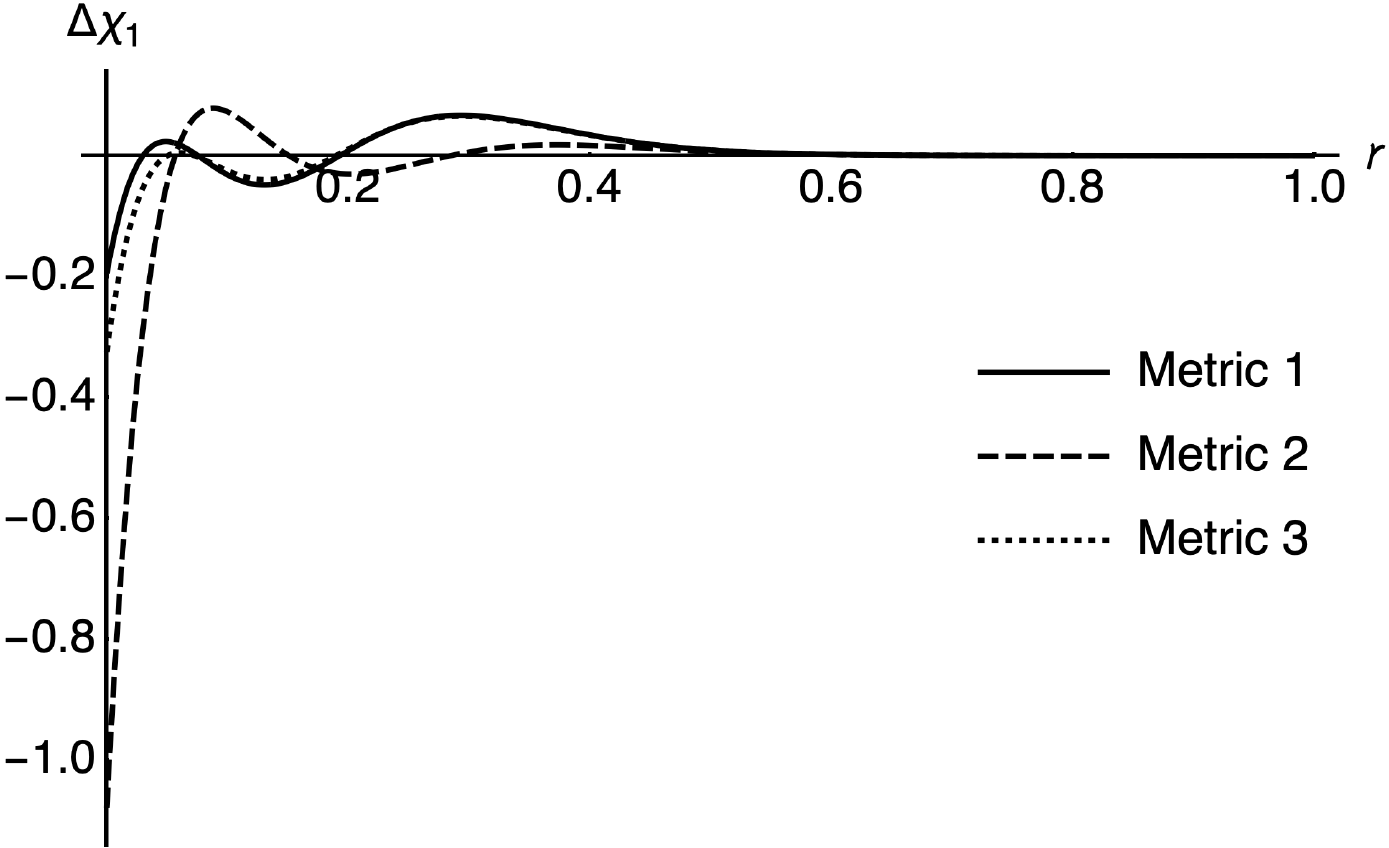}
        \caption{(b) $\Delta \chi_1$, $Z_\textrm{eff} = 16.43$, i.e. Chlorine}
    \end{subfigure}
    \vspace{2em}
    
        \begin{subfigure}[t]{0.45\textwidth}
        \centering
        \includegraphics[width=\textwidth]{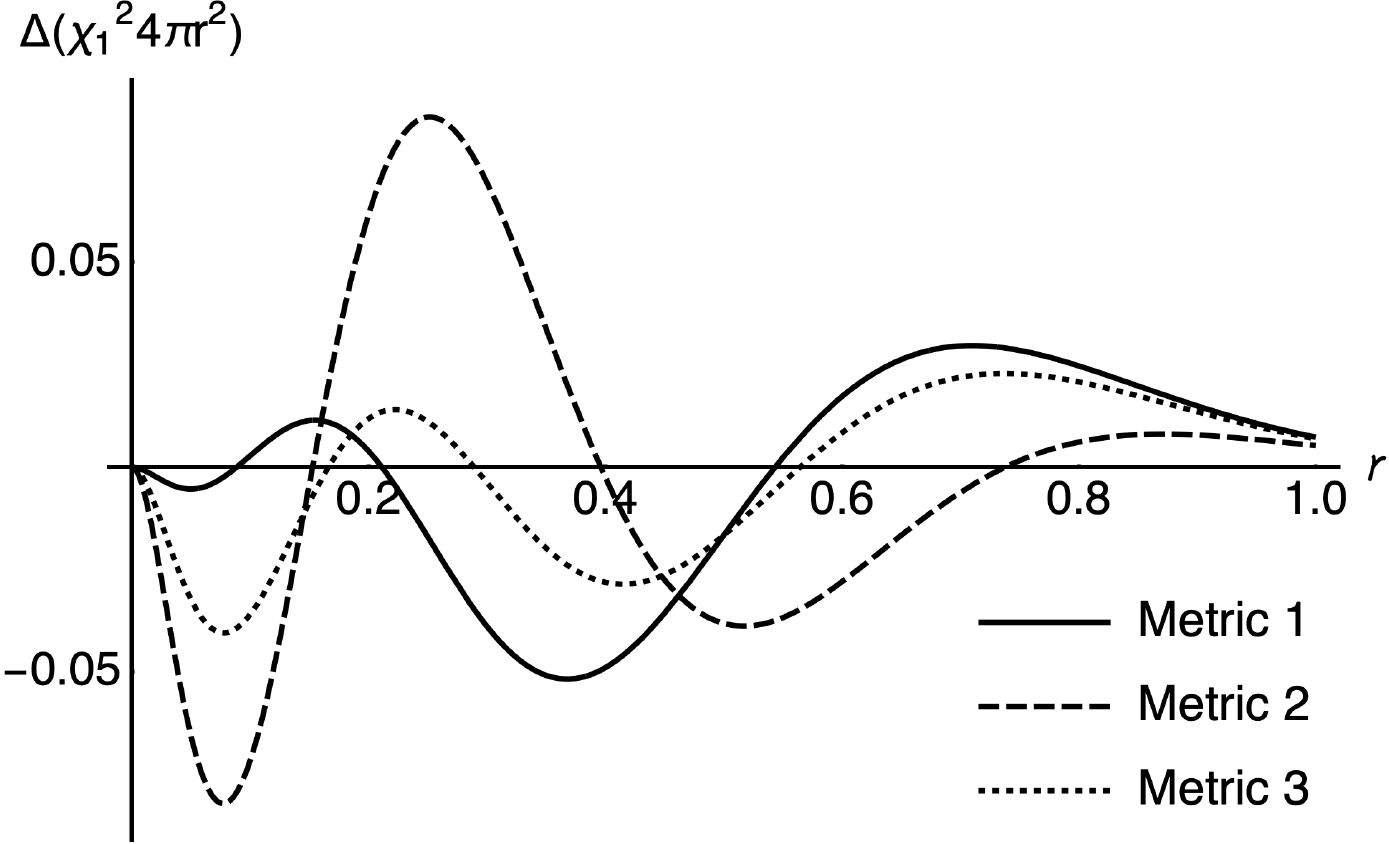}
        \caption{(c) $\Delta 4\pi r^2 (\chi_1^2)$,  $Z_\textrm{eff}  = 5.67$, i.e. Carbon}
    \end{subfigure}%
    \hspace{1em}
    \begin{subfigure}[t]{0.45\textwidth}
        \centering
        \includegraphics[width=\textwidth]{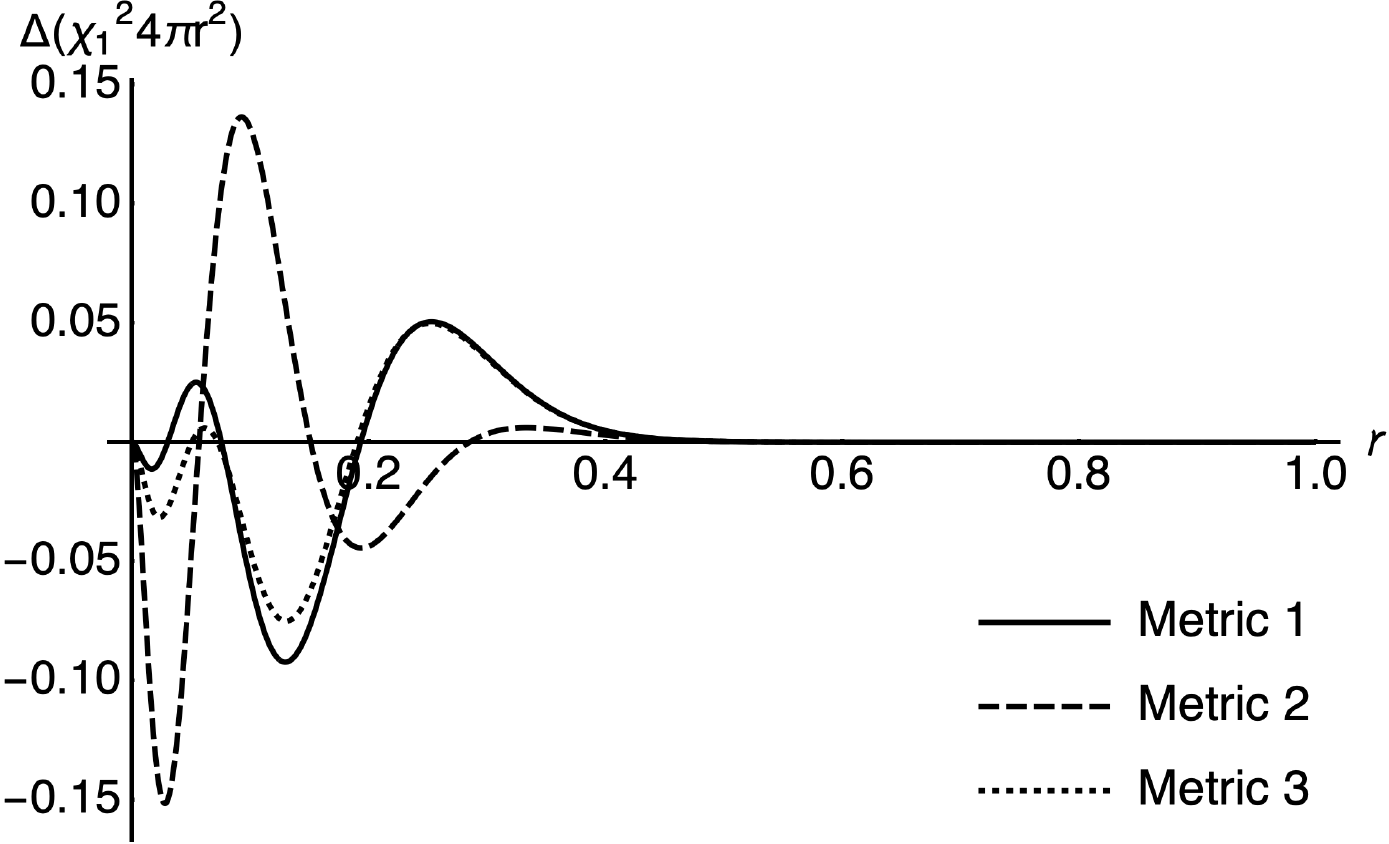}
        \caption{(d) $\Delta 4\pi r^2 (\chi_1^2)$, $Z_\textrm{eff} = 16.43$, i.e. Chlorine}
    \end{subfigure}
    \caption{Difference in the wavefunction (a, b) and probability density (c,d) between the original Slater function and a ramp-Gaussian function optimised using three different metrics (see text, section 2.1).}
    \label{fig:metricres}
\end{figure*}

\section{Fitting Slater Functions with Mixed Ramp-Gaussian Functions} \label{sec:Slater}

 %\clearpage
This section focuses on the creation of ramp and mixed ramp-Gaussian basis functions derived from the single-zeta Slater basis set for Li to Ne. The parameters for the derived basis functions are determined by minimising the difference between the $\mathcal{S}$ and $\mathcal{RG}$ core basis functions. The parameters are heavily dependent on the metric used in the fitting and hence we consider three potential metrics.

 \subsection{Fitting metric}
 When fitting one new basis function, A, against an existing basis function, B, there are three metrics we consider in this paper: 
 \begin{description} 
 \item[\textbf{Metric 1}] Minimising $\mathcal{L}_1 = \Braket{A^2 - B^2| A^2-B^2}$, as used for developing R-31G \cite{14McGiGi}; % =  \int (\mathcal{S}^2 - \mathcal{R}^2)^2 4 \pi r^2 dr =   \int (\mathcal{S}^4 + \mathcal{R}^4 - 2 \mathcal{S}^2\mathcal{R}^2) 4 \pi r^2 dr  $  
\item[\textbf{Metric 2}] Minimising $\mathcal{L}_2 = 1-|\Braket{A|B}|$, equivalent to minimising  $\Braket{A-B|A-B}$ as used for developing STO-3G \cite{sto-nG}; % =  \int (\mathcal{S} - \mathcal{R})^2 4 \pi r^2 dr =   \int (\mathcal{S}^2 - 2 \mathcal{S}\mathcal{R} + \mathcal{R}^2) 4 \pi r^2 dr  =   2 - 8\pi \int ( \mathcal{S}\mathcal{R})  r^2 dr  $. This is equivalent to maximising the overlap -  $\Braket{\mathcal{S}|\mathcal{R}} =  4 \pi \int \mathcal{S} \mathcal{R}  r^2 dr$.
\item[\textbf{Metric 3}] Minimising  $\mathcal{L}_3 = \int |A^2-B^2|  d\mathbf{r}$.
 \end{description}

 To investigate the characteristics of the fit preferred by each metric, we used each to fit a normalised mixed ramp-Gaussian function $c_1 S_n+c_2 s_\alpha$ to the Slater function $\mathbb{S}_{Z_\textrm{eff}}$. This was done for carbon (${Z_\textrm{eff} = 5.67}$) and chlorine (${Z_\textrm{eff} = 16.43}$). The fit parameters and several key basis function properties are shown in \Cref{tab:metrics}. \Cref{fig:metricres} visually shows the error in the wavefunction itself and the error in the probability density from mixed ramp-Gaussian approximations to the carbon (a,c) and chlorine (b,d) Slater basis functions. 
 
 For both carbon and chlorine, all metrics gave an optimal ramp degree $n= {\lceil}Z_\textrm{eff}{\rceil}+1$. However, the coefficients and Gaussian exponent varied between the three metrics with metrics 1 and 3 produced similar fit parameters, but metric 2 giving smaller Gaussian exponents with a slightly smaller contribution to the basis function. 
 
 We can compare the three fits by considering the value of all three metrics for each of the three fits, as shown in columns 6 to 8 of \Cref{tab:metrics}. It is evident that fitting to metrics 1 and 3 give comparable results while metric 2 is an outlier.  In particular, when the ramp Gaussian is fitted using metric 2, the $\mathcal{L}_1$ value is large. 
 
  in \Cref{tab:metrics} columns 9 to 11, we explore other properties of the core basis function to assess the quality of a particular fit. These properties are: 
 \begin{enumerate}
     \item $\chi(0)$, the value of the function at the origin
     \item $\frac{\chi^\prime(r)}{\chi(r)}\Big|_{r=0}$, the value of the cusp at the origin
     \item $\mathbb{E}(\chi_1) = \Braket{\chi_1|-\frac{Z_{\textrm{eff}}}{r} - \frac{\nabla^2}{2}|\chi_1}$, the energy of the hydrogenic atom with nuclear charge $Z_\textrm{eff}$ as calculated only using the tightest basis function, $\chi_1$.
 \end{enumerate}
 The signed error in these properties, calculated as the value based on the fitted ramp-Gaussian approximate function minus the Slater function, is given in \Cref{tab:metrics}. 
 
\Cref{tab:metrics} shows that, regardless of the choice of metric,  all fitted ramp-Gaussian functions have a smaller electron density at the nucleus and a slightly sharper cusp than the true Slater function. The fits obtained using metric 2 has higher errors in these core properties than metrics 1 and 3, with metric 1 fit giving slightly superior performance.  Further, as known from the variational principle, the approximate hydrogenic energy is higher than the exact energy, but there seems to be no meaningful pattern on which metric gives the lowest error in energy within the carbon and chlorine fits. 
 
 \Cref{fig:metricres} shows that fits from metric 2 performs consistently poorly at modelling both the wavefunction and probability density and produces the largest errors. Metric 3 is slightly superior at fitting probability density compared to metric 1, however the difference in performance becomes much smaller as $Z_\textrm{eff}$ increases.
 
 For describing properties of core electrons (as is the major goal for new mixed ramp-Gaussian basis sets), modelling the core region is the most important aspect when deriving the child basis functions. Hence, we select metric 1 as the way in which mixed ramp-Gaussian radial functions are fit to a given core basis function in this paper. These results additionally justifies the choice of this metric when defining the R-31G basis set \cite{14McGiGi}.

\begin{table*}
    \sisetup{round-mode=places,retain-explicit-plus}
    \caption{    \label{tab:singleSlaterRG}
Parameters of the derived Ramp-Gaussian basis function from the single-zeta Slater basis set for first row elements; definitions for each parameter are in the text.  } 
    \centering	
    \begin{tabular}{lccS[round-precision=7,table-parse-only]cS[round-precision=7,table-parse-only]S[round-precision=7,table-parse-only]S[round-precision=7,table-parse-only]cS[round-precision=7,table-parse-only]S[round-precision=7,table-parse-only]S[round-precision=7,table-parse-only]S[round-precision=7,table-parse-only]S[round-precision=7,table-parse-only]cS[round-precision=7,table-parse-only]S[round-precision=7,table-parse-only]S[round-precision=7,table-parse-only]S[round-precision=7,table-parse-only]S[round-precision=7,table-parse-only]S[round-precision=7,table-parse-only]S[round-precision=7,table-parse-only]cS[round-precision=7,table-parse-only]S[round-precision=7]}
    \toprule
         & \mc{1}{c}{STO}  & \mc{1}{c}{STO-R } & \mc{4}{c}{STO-RG } & \mc{6}{c}{STO-R2G }\\% & \mc{8}{c}{STO-3G  \alert{(OPTIONAL)}} \\
        \cmidrule(r){2-2} \cmidrule(r){3-3} \cmidrule(r){4-7} \cmidrule(r){8-13}% \cmidrule(r){14-21}
        &\mc{1}{c}{$Z_\textrm{eff}$}   &   \mc{1}{c}{$n$} & \mc{1}{c}{$c_1$} & \mc{1}{c}{$n$} & \mc{1}{c}{$c_2$} & \mc{1}{c}{$\alpha_1$} & \mc{1}{c}{$c_1$} & \mc{1}{c}{$n$} & \mc{1}{c}{$c_2$} & \mc{1}{c}{$\alpha_1$} &  \mc{1}{c}{$c_3$} & \mc{1}{c}{$\alpha_2$} \\
         \midrule
Li & 2.69 & 1& 0.2730262183 & 4 & 0.8163954599 & 1.526017431 & 0.26100294548626153723 & 4 & 0.6241182081325588286 & 1.9483467430956981748 & 0.2493601925789782278 & 0.5095018358339954903\\
Be & 3.68 & 2& 0.3622450889 & 5 & 0.7272642323 & 2.463572762 &  0.35284422463997312946 & 5 & 0.5377595879374242024 & 3.1099060921013296476 & 0.2286710484186325175 & 1.0311567175530396492\\
B & 4.68 & 3& 0.4360306876 & 6 & 0.6492041387 & 3.587269047 & 0.4268055413810403992 & 6 & 0.3706169766682207794 & 4.9726802052525927490 & 0.3077689271822700753 & 2.111790678537136899  \\
C & 5.67 & 4& 0.4938827304 & 7 & 0.5856474836 & 4.896756543 & 0.4840874796299068085 & 7 & 0.2280053055222888326 & 7.827448505540135692 & 0.3833578821081713184 & 3.441898024128910706 \\
N & 6.67 & 5& 0.5427120233 & 8 & 0.5310134749 & 6.401889787 & 0.5327119049390098500 & 8 & 0.1608235084236976905 & 11.343410766425142666 & 0.3944489440859021030 & 4.787906251009735353 \\
O & 7.66 & 6& 0.5816851812 & 9 & 0.4866186707 & 8.101958121 & 0.5714925314201395627 & 9 & 0.1263555458421248143 & 15.487827813631293927 & 0.3839337309447161684 & 6.217456652498315880 \\
F & 8.65 & 7& 0.6144117594 & 10 & 0.4489925087 & 10.00106287 & 0.604182464156271328 & 10 & 0.106602613215013402 & 20.14928473488953853 & 0.365600715832689695 & 7.75845463419095736\\
Ne & 9.64 & 8 & 0.642248393 & 11 & 0.4167850089 & 12.10057083 & 0.632081458884322032 & 11 & 0.0936413484585769147 & 25.329387459902001776 & 0.345900580714023483 & 9.430019885412059985 \\
         \bottomrule
    \end{tabular}
\end{table*}

 \subsection{Creation of the derived STO-R$n$G basis sets}
  
 The core STO-R basis functions are formed by fitting the core Slater basis function with a single ramp. Using metric 1 and allowing only integer ramp degrees, a single ramp function best fit a Slater function with exponent $Z_\textrm{eff}$ if the ramp degree was $n= \lceil Z_\textrm{eff}\rceil -2$, where $\lceil x \rceil$ is the Ceiling function of $x$, which gives the smallest integer greater than or equal to the input. 
 
 The core basis functions in the STO-RG basis set have the form $\chi_{1}^\textrm{STO-RG} = {c_1}S_n + {c_2}s_{\alpha_1}$ while STO-R2G core basis functions have the form $\chi_{1}^\textrm{STO-R2G} = {c_1}S_n + {c_2}s_{\alpha_1} + {c_2}s_{\alpha_2}$. The adjustable parameters, $c_1, c_2, n, \alpha_1, \alpha_2$ are determined by minimising the difference between the normalised $\chi_1$ STO-R$n$G basis function and the STO basis function for that element, as quantified using metric 1. This was done in Wolfram Mathematica \cite{wolfram} using the \texttt{FindMinimum} and \texttt{NIntegrate} functions. The optimisations were very sensitive to the initial guess for the parameters, especially for the larger STO-R2G basis set. Note that one of the coefficients was not fit in the numerical optimisation, but rather constrained by the normalisation of the basis function. Further, the Mathematica command for a global minimisation, \texttt{NMinimize}, was tested but found to be less reliable than the local minimisation command, \texttt{FindMinimum}, with a good initial guess.  
 
 The optimized parameters of these basis functions are given in \Cref{tab:singleSlaterRG}. The robustness of optimisation was tested manually for all basis functions through using different initial guesses and  grid searches when necessary. A ramp degree of $n = {\lceil}Z_\textrm{eff}{\rceil }+ 1$ was found to give the best fit for the ramp+Gaussian fit, significantly more than the equivalent value when fitting with a single ramp.  \Cref{tab:singleSlaterRG} shows a smooth increase in the $\alpha$ values as $Z_\textrm{eff}$ increases, as a tighter Gaussian function is needed to match the increasing contraction of the Slater function. The Gaussian exponents and Slater exponent are approximately quadratically related, i.e. $\alpha \propto Z_\textrm{eff}^2$. 

It is illustrative to compare the Gaussian exponents of STO-RG and STO-R2G to the exponents in the STO-$n$G basis sets (sourced from Basis Set Exchange). For STO-RG, the Gaussian exponents are generally between the smallest and second smallest exponent of STO-$n$G, indicating the Gaussian in STO-RG is modelling the valence region, as expected. Along the periodic table from Li to Ne, the ratio between the exponent of STO-RG and the smallest Gaussian exponent of STO-$n$G becomes smaller as does the relative contribution of the ramp function to the overall core basis function. These behaviours occur because, with a finite range of 0 to 1 Bohr, the ramp is less able to model the diffuse Li $1s$ orbitals than the tighter Ne $1s$ orbitals.  Similar behaviour is observed for STO-R2G, with the smaller exponent of STO-R2G is generally between the smallest and second smallest exponent in STO-$n$G, and the larger exponent of STO-R2G between the second and third smallest exponent.

\begin{figure}
  \centering
   \includegraphics[width=0.5\textwidth]{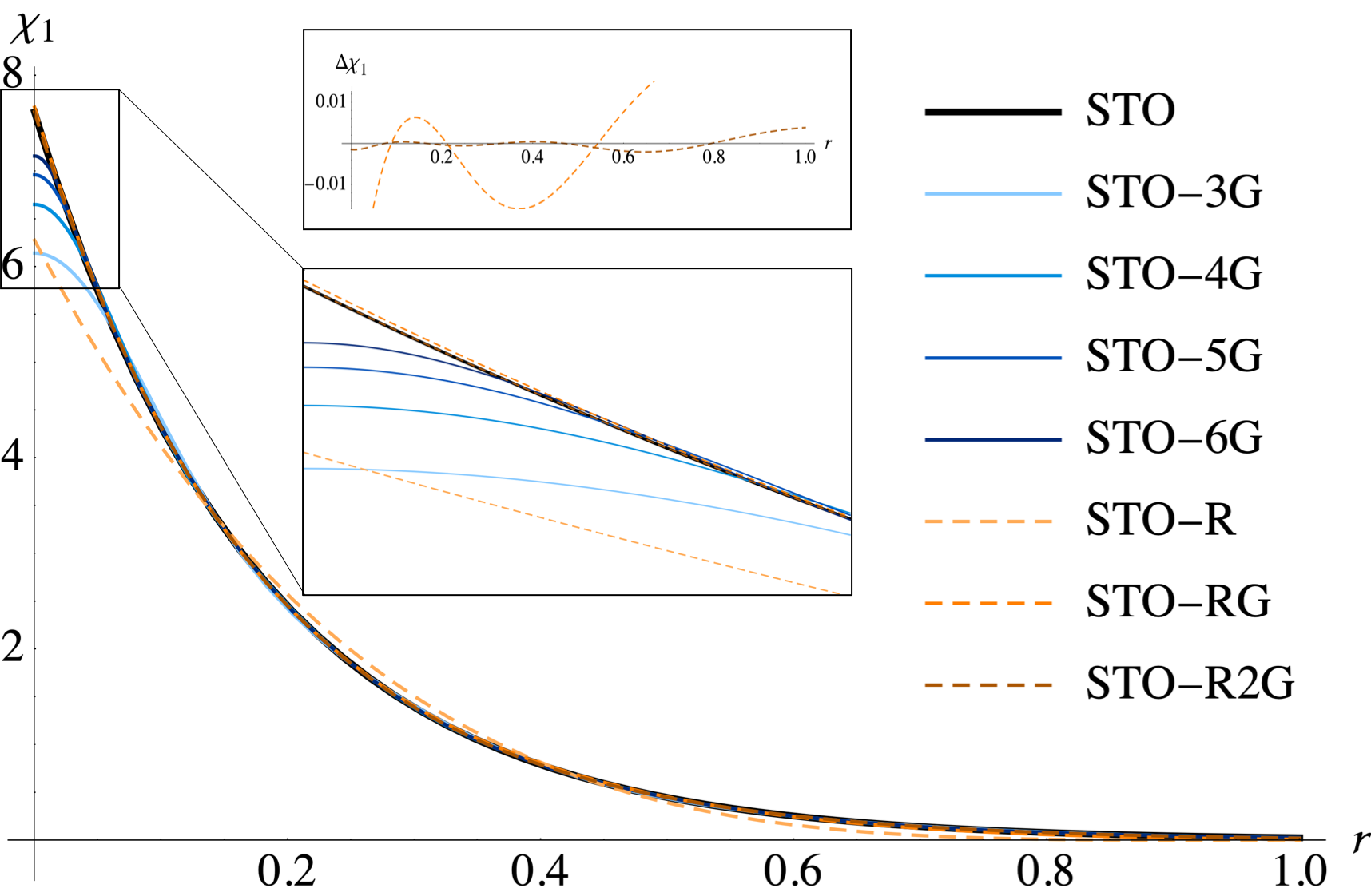}
    \caption{The wavefunction for the STO parent function, and the STO-$n$G and STO-R$n$G child basis functions for carbon. The top inset is a plot of the difference between STO and STO-RG, and between STO and STO-R2G. The bottom inset is a magnified view of the near core region, between 0 and 0.05 Bohr. } 
\label{fig:wavefunction}
\end{figure}

\begin{figure}
  \centering
   \includegraphics[width=0.5\textwidth]{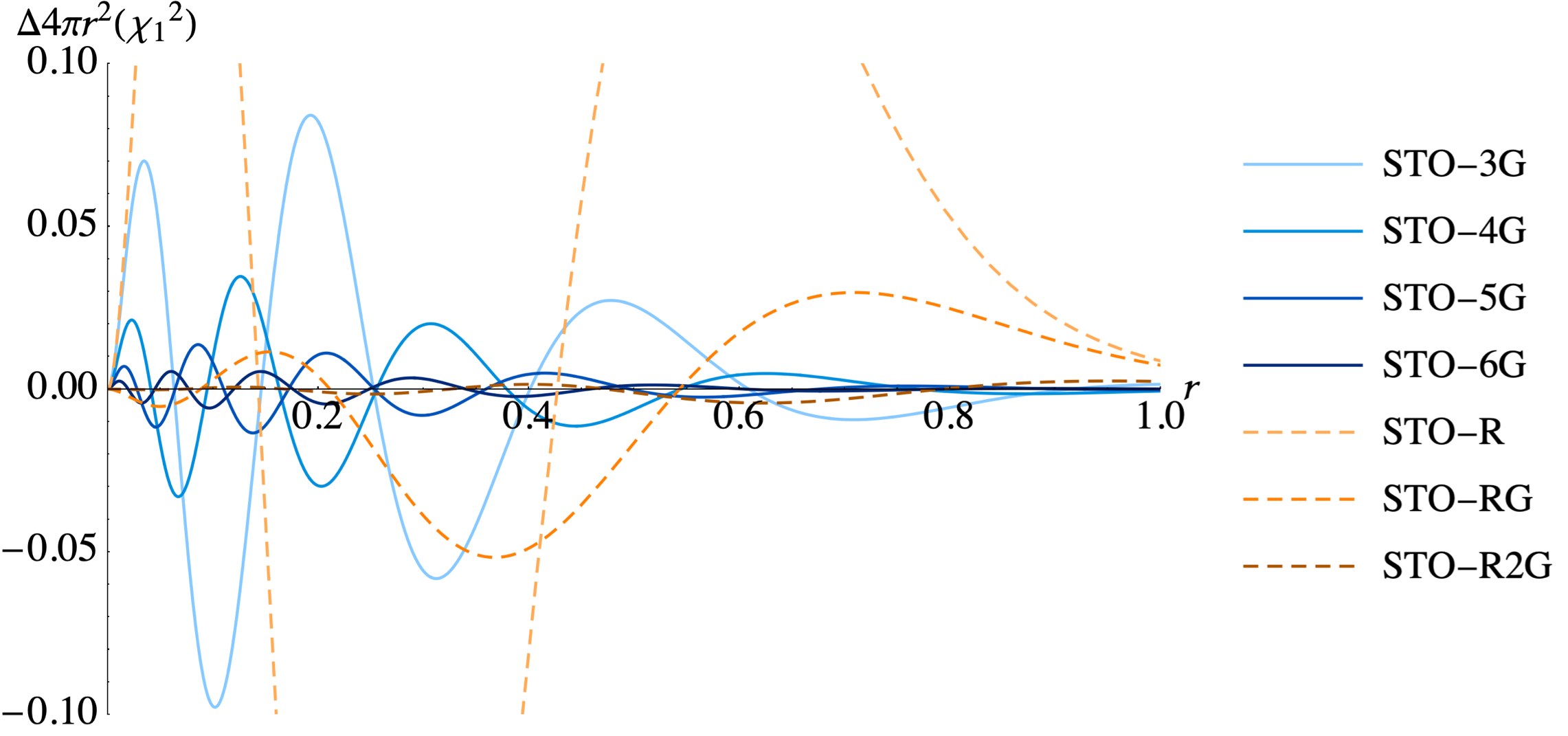}
    \caption{The difference in the probability density for the STO parent function, and the STO-$n$G and STO-R$n$G child basis functions for carbon.  
} 
\label{fig:probdensity}
\end{figure}

 \subsection{Evaluation of the STO-R$n$G basis sets}

A visual representation of the different derived STO-R$n$G and STO-$n$G basis functions compared to the parent STO basis set for carbon can be seen in \Cref{fig:wavefunction}. The improved modelling of the electron-nuclear cusp with STO-R$n$G basis sets over the STO-$n$G basis sets is clear. The error of the STO-$n$G and STO-R$n$G basis functions against the STO parent basis set decreases as $n$ increases, as expected. The STO-R is visually a poor fit, with substantial improvements seen with the addition of one then two Gaussians.  

\Cref{fig:probdensity} is the difference in the probability density for the various core basis functions for carbon compared to the parent STO basis function. The STO-R basis set performs poorly. STO-RG visually appears superior to the STO-3G basis set but poorer than the STO-4G basis set over this domain. The STO-R2G basis set outperforms the STO-6G basis set. Overall, the STO-$n$G all-Gaussian basis sets have larger errors in the inner core ($r < 0.3$) but smaller errors in the outer core region compared to the STO-R$n$G ramp-Gaussian core functions.

 \begin{table*}
    \sisetup{round-mode=places,retain-explicit-plus}
    \caption{    \label{tab:fitqualitymetric1}
Value of fitting metrics $\mathcal{L}_1$ and $\mathcal{L}_2$ for STO-$n$G and STO-R$n$G core $\chi_1$ basis function compared to the parent Slater basis function.}
    \centering
    \begin{tabular}{lcS[round-precision=3]S[round-precision=3]S[round-precision=3]S[round-precision=3]S[round-precision=3]S[round-precision=3]S[round-precision=3]S[round-precision=3]}

    \toprule
    & $Z_\textrm{eff}$ & \mc{1}{c}{STO-3G} & \mc{1}{c}{STO-4G} & \mc{1}{c}{STO-5G} & \mc{1}{c}{STO-6G} & \mc{1}{c}{STO-R} & \mc{1}{c}{STO-RG} & \mc{1}{c}{STO-R2G} \\
    \midrule
        \mc{8}{l}{\textbf{STO-R$n$G fitting metric:} $1000 \mathcal{L}_1 = 1000\braket{\chi_1^2-\mathbb{S}_{Z_\textrm{eff}}^2|\chi_1^2-\mathbb{S}_{Z_\textrm{eff}}^2}$} \\
%\cmidrule(r){1-4}
    \vspace{-0.7em}\\
Li & 2.69 & 1.9676451067166092 & 0.2998761492551743 & 0.05225853172494843 & 0.010174027272574269 & 82.96933021177144 & 0.255904 & 0.0049926503648724780497 \\
Be & 3.68 & 5.0377125793775 & 0.7677654087483601 & 0.13379621244826567 & 0.026048307722760042 & 95.1446874912674  & 0.226715 & 0.00018035802669567776117\\
B & 4.68 & 10.361615885817399 & 1.5791473093430273 & 0.27519334216522795 & 0.053576459306205325 & 113.2926537405221 & 0.272651 & 0.00043130681681281756838 \\
C & 5.67 & 18.426341139549113 & 2.808240247897742 & 0.48938375300237075 & 0.0952763748368511 & 128.2205866560976 & 0.375024 & 0.0011887130316742352327 \\
N & 6.67 & 29.99628223373234 & 4.571540663383699 & 0.7966689023262026 & 0.15510062922044265 & 147.1949676907123  & 0.496676 & 0.0016017468625203396250\\
O & 7.66 & 45.43350457664651 & 6.924228547770312 & 1.206664835856861 & 0.23492128460231554 & 160.87382529513772 & 0.656145 &0.002116963124 \\
F & 8.65 & 65.42417467115874 & 9.970878117309534 & 1.7375955096537328 & 0.3382862768177908 & 173.52801470190693 & 0.837439 & 0.002716437259 \\
Ne & 9.64 & 90.55679245345158 & 13.801178949072865 & 2.4050907375178667 & 0.46823854265424397 & 185.17705536591444 & 1.03782 & 0.0034346053785640955188 \\
\vspace{-0.5em} \\

        \mc{8}{l}{\textbf{STO-$n$G fitting metric:} $1000 \mathcal{L}_2 = 1000(1-\braket{\chi_1|\mathbb{S}_{Z_\textrm{eff}}^2})$} \\
%        \cmidrule(r){1-4}
        \vspace{-0.7em} \\
Li & 2.69 & 1.6526373498337854 & 0.21881416802149722 & 0.03441962524441955 & 0.0061851908550458745 & 68.78815267 & 3.829880645 & 0.1641621207\\
Be & 3.68 & 1.652637478454233 & 0.21881324953731962 & 0.0344202054025633 & 0.006186036172195486 & 30.53895464 & 1.918406634 & 0.01071372315\\
B & 4.68 & 1.6526373956382567 & 0.21881323056915924 & 0.03442022489807961 & 0.006084269580375334 &  16.7190424 & 1.187543712 & 0.02740283065\\
C & 5.67 & 1.6526374884029416 & 0.21881325018902054 & 0.03442023818855944 & 0.006186025730547939 &  10.556331 & 0.8866113596 & 0.05484085832\\
N & 6.67 & 1.652637459343964 & 0.21881326170092308 & 0.034420196074469445 & 0.006186044724243445 & 7.189767091 & 0.720866305 & 0.05257031867 \\
O & 7.66 & 1.6526374895842189 & 0.21881323540640096 & 0.03442021138333473 & 0.006186049559264717 & 5.259467252 & 0.6322300491 &  0.04675999816 \\ 
F & 8.65 & 1.652639399565281 & 0.21881392366251973 & 0.03441727263853167 & 0.006184449875545894 & 4.022642591 & 0.5718331373 & 0.0412119473 \\
Ne & 9.64 & 1.6526372629543928 & 0.21881378154176012 & 0.03442068903569684 & 0.006185755037080298 & 3.183078759 & 0.5259304286 & 0.03693269398 \\
\bottomrule
    \end{tabular}
    \end{table*}
    
\Cref{tab:fitqualitymetric1} details the $\mathcal{L}_1$ and $\mathcal{L}_2$ residuals of the STO-$n$G and STO-R$n$G basis functions when compared to their parent Slater basis function. 

When using metric 1 (the metric used to fit the STO-R$n$G basis functions), the STO-R basis functions are much worse than STO-3G, STO-RG is approximately equivalent to STO-4G or STO-5G and  STO-R2G greatly outperforms even the STO-6G basis function. Using metric 1 to compare fit quality for different elements is problematic. We expected the fit for lithium to be the poorest, due to the issue with the Slater function extending beyond the domain of the ramp function. However, the $\mathcal{L}_1$ metric is very sensitive to the overall size of the basis function with more contracted basis functions yielding larger values. Therefore, in all cases $\mathcal{L}_1$ increases with nuclear charge but this cannot be interpreted as a representing poorer overall fit quality. Nevertheless, it is notable that, w grthile the STO-$n$G basis sets have a rapid increase in the $\mathcal{L}_1$ residual as $Z_\textrm{eff}$ increases, the $\mathcal{L}_1$ residual STO-R$n$G basis functions are much more stable for larger $Z_\textrm{eff}$. This result implies that the quality of the STO-$n$G fit in the core region becomes poorer for higher nuclear charge.  

The latter half of \Cref{tab:fitqualitymetric1} shows the residuals when using metric 2 (the metric used to fit the STO-$n$G basis functions). With this $\mathcal{L}_2$ metric, the residuals of the STO-$n$G basis functions are independent of the element, while the STO-R$n$G basis function residual decrease as $Z_\textrm{eff}$ increases. STO-R2G is slightly worse than STO-5G quality according to this metric. %Detailed investigations demonstrate that this is because metric 1 prioritises the core region while metric 2 focuses on the valence region. %This is possibly because metric 1 prioritises the core region, while metric 2 focuses on the valence region.
  
 \begin{table*}
    \sisetup{round-mode=places,retain-explicit-plus}
    \caption{    \label{tab:valueatnuclei}
Value of key properties for the core $\chi_1s$ basis function of the minimal STO basis set,  and the errors obtained using STO-$n$G and STO-R$n$G approximations to these core $\chi_1$ basis functions (Error = Approximate value - Exact value). }
    \centering
    \begin{tabular}{lcS[round-precision=3]S[round-precision=3]S[round-precision=3]S[round-precision=3]S[round-precision=3]S[round-precision=3]S[round-precision=3]S[round-precision=3]}

    \toprule
    & $Z_\textrm{eff}$ & \mc{1}{c}{STO} & \mc{1}{c}{$\Delta$ STO-3G} & \mc{1}{c}{$\Delta$ STO-4G} & \mc{1}{c}{$\Delta$ STO-5G} & \mc{1}{c}{$\Delta$ STO-6G} & \mc{1}{c}{$\Delta$ STO-R} & \mc{1}{c}{$\Delta$ STO-RG} & \mc{1}{c}{$\Delta$ STO-R2G} \\
    \toprule
         \mc{8}{l}{\textbf{Value at nuclei:} $\chi_1(r=0)$} \\
%        \cmidrule(r){1-4}
        \vspace{-0.7em} \\  
Li & 2.69 & 2.4891638416032555 & -0.48179937976336573 & -0.3158062087720217 & -0.21476491914753337 & -0.15030666571064355 & -0.9440670335104973 & +0.0232845 & -0.01016941269 \\
Be & 3.68 & 3.9828760555372233 & -0.7709204113222707 & -0.5053170732534196 & -0.34364232684620655 & -0.2405035823528192 & -1.0922646128966695 & +0.0296015 & -0.002998873426 \\
B & 4.68 & 5.712074238960018 & -1.1056218067268349 & -0.724704609379577 & -0.49283745100085863 & -0.34492139067275307 & -1.2339582478786335 & +0.0383929 & +0.00004293117748  \\
C & 5.67 & 7.6172753008294345 & -1.4743900960471805 & -0.9664220551117211 & -0.657218094543981 & -0.4599646014403662 & -1.3410623969164304 & +0.0493523 & +0.001446150519 \\
N & 6.67 & 9.71882617184804 & -1.8811636026834702 & -1.2330508718676771 & -0.8385397875293776 & -0.5868654904149793 & -1.4558079226594831 & +0.058767 & +0.001902714045 \\
O & 7.66 & 11.961020042307625 & -2.3151597888903375 & -1.5175234059434768 & -1.031996150337232 & -0.7222590233512332 & -1.5387722684241734 & +0.0694724 & +0.002828242255 \\
F & 8.65 & 14.35321614204408 & -2.7781902193695913 & -1.8210270827506232 & -1.2383946953566145 & -0.8667103500296633 & -1.6120214589017934 & +0.0799398 & +0.003821934959 \\
Ne & 9.64 & 16.88653772232112 & -3.268536711733317 & -2.142435692298381 & -1.456969546579991 & -1.0196834529155598 & -1.676945032703495 & +0.090181 & +0.004885179033 \\
    \vspace{-0.5em}\\

          \mc{8}{l}{\textbf{One-electron energy:} $\mathbb{E}(\chi_1) = \Braket{\chi_1|-\frac{Z_{\textrm{eff}}}{r} - \frac{\nabla^2}{2}|\chi_1}$} \\
%        \cmidrule(r){1-4}
        \vspace{-0.7em} \\  
        Li & 2.69 & -3.6180499999999998 & 0.03685278657711821 & +0.010992849640417024 & +0.0035776188143752385 & +0.001253024454708651 & +1.8930499999999983  & +0.03715858832 & +0.002409218925 \\
Be & 3.68 & -6.7712 & +0.0689701885692724 & +0.020573176035138196 & +0.006695531456374582 & +0.0023450442565877694 & +0.8911999999999951 & +0.03125436441 & +0.0003145788607 \\
B & 4.68 & -10.951199999999998 & +0.11154689406516027 & +0.03327341765689873 & +0.010828819783101906 & +0.0037979099504177327 & +0.6911999999999949 & +0.02951329496 & +0.0006702198628  \\
C & 5.67 & -16.07445 & +0.16373136923493448 & +0.04883981819264349 & +0.015894756469556626 & +0.005567003517487734 & +0.6037357142856976  & +0.03160316627 & +0.002069741219 \\
N & 6.67 & -22.24445 & +0.2265778457154397 & +0.06758625421404929 & +0.021995867819502024 & +0.007703893619186886 & +0.5561166666666573  & +0.03520278844 & +0.002708590336 \\
O & 7.66 & -29.3378 & +0.29882939937894903 & +0.08913808718582317 & +0.02900993285764386 & +0.010160533824013385 & +0.5241636363636175  & +0.0405370117 & +0.003129381396 \\
F & 8.65 & -37.41125 & +0.38106405962244594 & +0.11366791259679587 & +0.03699316106803252 & +0.012956457722175685 & +0.5016346153845888  & +0.04656308313 & +0.003477688069 \\
Ne & 9.64 & -46.464800000000004 & +0.47328184187779954 & +0.14117564322809528 & +0.04594556086592405 & +0.016091977636669696 & +0.4847999999999857  & +0.05295397305 & +0.003841721004  \\
\bottomrule

    \end{tabular}
\end{table*}

 \begin{table}
    \sisetup{round-mode=places,retain-explicit-plus}
    \caption{    \label{tab:electronnuclearcusp}
Electron-nuclear cusp evaluated as $\chi^\prime(r)/\chi(r)$ at $r=0$ for the core $\chi_{1s}$ basis function of the minimal STO basis set and the STO-R$n$G approximations to these functions. Note that STO-$n$G core functions all have $\chi^\prime(r)=0$ and are thus not included in this table. }
    \centering
    \begin{tabular}{lcS[round-precision=3]S[round-precision=3]S[round-precision=3]S[round-precision=3]}

    \toprule
    & $Z_\textrm{eff}$ & \mc{1}{c}{STO} & \mc{1}{c}{$\Delta$ STO-R} & \mc{1}{c}{$\Delta$ STO-RG} & \mc{1}{c}{$\Delta$ STO-R2G} \\
    \toprule
           \mc{5}{l}{\textbf{Electron-nuclear cusp:} $\chi^\prime(r)/\chi(r)$ at $r=0$} \\
%        \cmidrule(r){1-4}
        \vspace{-0.7em} \\ 
Li & 2.69 & -2.69 & +1.69 & -0.038128781 & +0.046640566 \\
Be & 3.68 & -3.68 & +1.68 & -0.049912172 & +0.017125326 \\
B & 4.68 & -4.68 & +1.68 & -0.061618113 & +0.007540065 \\
C & 5.67 & -5.67&+1.67 & -0.075497797 & +0.003042563 \\
N & 6.67 & -6.67 & +1.67 & -0.083751063 & +0.001914765 \\
O & 7.66 & -7.66 &+1.66 & -0.094050439 & -0.0006158974606 \\
F & 8.65 & -8.65 &+1.65 & -0.103126198 & -0.003030606737 \\
Ne & 9.64 & -9.64 & +1.64 & -0.111278737 & -0.00537494\\
\bottomrule
\end{tabular}
\end{table}

%\clearpage 
The top part of \Cref{tab:valueatnuclei} shows the value at the core basis function at the nucleus for the STO core basis function with exponent $Z_\textrm{eff}$, and the error of this value for the STO-$n$G and STO-R$n$G basis sets. For the value at the nuclei, only STO-RG is needed to outperform STO-6G, and the discrepancy for STO-R2G is almost zero. For STO-R and STO-RG , the error becomes more negative as $Z_\textrm{eff}$ increases; this indicates the ramp-Gaussian is progressively overshoots the Slater function at the nucleus more as $Z_\textrm{eff}$ increases. STO-R2G improves the STO-RG results for this metric by about an order of magnitude. For the STO-R2G, the ramp-Gaussian overshoots and then undershoots the true value at the nucleus, with a minimum discrepancy at boron.

The bottom part of \Cref{tab:valueatnuclei} shows the energy of a hydrogenic atom with nuclear charge $Z_\textrm{eff}$ as calculated with the Slater function (i.e. the exact answer) compared to the core STO-$n$G and STO-R$n$G basis function (not the full basis set).  The quality of the energetic description of the STO basis function by STO-R is extremely poor for all elements, but improves as $Z_\textrm{eff}$ increases. Unlike for the value at the nuclei, the STO-RG basis set is on par with STO-3G and STO-4G in terms of quality. STO-R2G is comparable with STO-6G for lithium and outperforms it for heavier elements. This error for the STO-R$n$G basis sets depends much less on effective nuclear charge than the comparable error for the STO-$n$G basis sets. 

\begin{figure}
  \centering
   \includegraphics[width=0.5\textwidth]{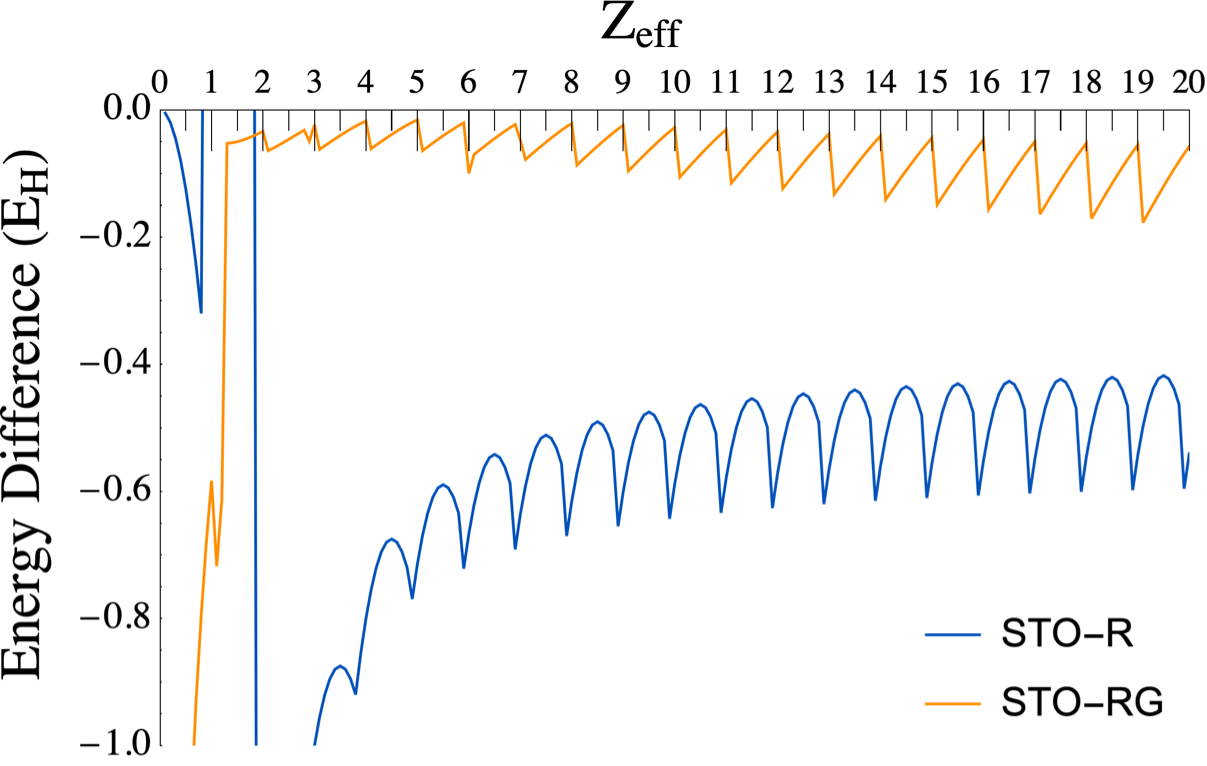}
    \caption{ Error in the energy of the hydrogenic ion with nuclear charge $Z_\textrm{eff}$ as calculated with a single ramp function with $n= \lceil Z_\textrm{eff} \rceil -2$ as well as a ramp Gaussian with $n= \lceil Z_\textrm{eff} \rceil + 1$. 
    %The outlier spikes in the STO-RG plot are due to optimisations that did not reach the true minimum, and those points should be disregarded. 
    STO-R2G optimisation is too sensitive to initial parameter guesses to generate a optimized basis function for each value of $Z_\textrm{eff}$ in this parameter space.} 
\label{fig:energydiffplot}
\end{figure}

 \Cref{fig:energydiffplot} is a visual demonstration and extension of the data presented in the latter half of \Cref{tab:valueatnuclei}. This figure shows that for STO-R, $Z_\textrm{eff}$ values below 3 generate large and unstable errors. This indicates that a single ramp function to describe the $1s$ orbital for H, He and Li would not be useful. This error is stabilised for STO-RG, indicating that Li can be successfully fitted with the addition of a Gaussian function. Oscillatory behaviour is visible for both STO-R and STO-RG with a period of one $Z_\textrm{eff}$ value. While the energy difference for STO-RG slowly increases with $Z_\textrm{eff}$ (in agreement with \Cref{tab:valueatnuclei}), for large nuclear charges the error in the energy overall for STO-R is seen to slightly decrease to an average asymptotic value of ~0.45 $E_\textrm{H}$, despite the quadratic increase in overall energy of the system with increasing $Z_\textrm{eff}$.
 
\Cref{tab:electronnuclearcusp} show the value of the electron-nuclear cusp at the nucleus for elements Li to Ne for the STO core basis function with exponent $Z_\textrm{eff}$ and the deviation from this value for the  STO-R, STO-RG and STO-R2G core basis functions. Note that Gaussians will have cusps of zero in all cases. The electron-nuclear cusp values in \Cref{tab:electronnuclearcusp} show a close agreement between the Slater and ramp-Gaussian basis function for STO-RG and even closer agreement for the STO-R2G basis functions. 

 %The stability of STO-R for high $Z_\textrm{eff}$ suggest the potential usefulness of fitting the innermost core basis function of second-row or higher elements with just a single ramp, especially if augmented by flexibility with other basis functions. This will be explored in future work.

When considering all the different measures of fit quality, it is evident that STO-R is a poor fit, STO-RG is a good fit approximately of STO-4G to STO-5G quality and STO-R2G is an exemplary fit better than STO-6G quality that considerably improves on the STO-RG fit. 

The improvement of the fit quality with the addition of a second Gaussian is notable for influencing the composition of future mixed ramp-Gaussian basis sets. Initially when developing these new types of basis functions, the major goal was faster calculations through reduction in the number of primitive functions within the core basis function.  However, with the focus of future development of mixed ramp-Gaussian basis sets now shifted to accurate description of core electrons, the additional Gaussian primitive in the core is likely to be an acceptable increase in computational time if the improved core electron description described in this section can be demonstrated for practical calculations.

%\clearpage

%\newpage

\section{Rampifying all-Gaussian core basis functions} \label{sec:Gauss} 
The focus of this section is to consider the generality of the results in \Cref{sec:Slater} and their implications on future development of mixed ramp-Gaussian basis sets derived from parent all-Gaussian basis sets. Our results in this section focus on carbon as an illustrative example of first-row element behaviour (Li-Ne).

\subsection{R-31G basis set}

\begin{table}
    \sisetup{round-mode=places,retain-explicit-plus}
    \caption{\label{tab:rn31gparameters} Fit parameters for the core R$n$-31G basis functions for carbon. Additionally, for ease of comparison, the  core STO-R$n$G basis functions for carbon have been reproduced along with the specifications for the two primitive Gaussians with the smallest exponents in the core basis function of 6-31G. All values have been truncated to 3 decimal places for convenient comparisons. }
    \centering
    
    \resizebox{0.48\textwidth}{!}{%
\begin{tabular}{lS[round-precision=3]S[round-precision=3]S[round-precision=3]S[round-precision=3]S[round-precision=3]S[round-precision=3]S[round-precision=3]}
 \toprule
 & \mc{1}{c}{$c_1$} & $n$ & \mc{1}{c}{$c_2$} & \mc{1}{c}{$\alpha_1$} & \mc{1}{c}{$c_3$} & \mc{1}{c}{$\alpha_2$} \\ \midrule
 \mc{3}{l}{6-31G \emph{(Two smallest exponents)}}   & 0.4679413484 &  9.286662960 & 0.3623119853 & 3.163926960 \\
R-31G & 0.5120939 & 7 & 0.5749880 & 4.545 &  &  \\
R2-31G & 0.50942046687621610399 & 7 & 0.5309957847380515936 & 4.9114501630376451996 & 0.07213968827154599653 & 0.887386389688618249  \\
STO-RG & 0.4938827304 & 7 & 0.5856474836 & 4.896756543  &   &  \\
STO-R2G & 0.4840874796299068085 & 7 & 0.2280053055222888326 & 7.827448505540135692 & 0.3833578821081713184 & 3.441898024128910706 \\ \bottomrule
\end{tabular}%
}
\end{table}

\begin{figure}
  \centering
   \includegraphics[width=0.5\textwidth]{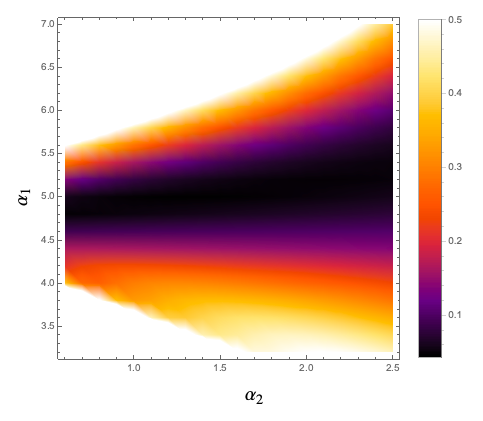}
    \caption{The residual ($1000 \mathcal{L}_1$) was calculated for the fit of the R2-31G across a parameter space of the exponents.
} 
\label{fig:r231gl1}
\end{figure}

\begin{table}
    \sisetup{round-mode=places,retain-explicit-plus}
    \caption{    \label{tab:r31gl1} $1000 \mathcal{L}_1$ between two core basis functions for carbon  in the specified basis set.  }
    \centering
    \resizebox{0.48\textwidth}{!}{%
    \begin{tabular}{lS[round-precision=3]S[round-precision=3]S[round-precision=3]S[round-precision=3]S[round-precision=3]S[round-precision=3]}
\toprule
 & \mc{1}{c}{STO} & \mc{1}{c}{STO-RG} & \mc{1}{c}{STO-R2G} & \mc{1}{c}{6-31G} & \mc{1}{c}{R-31G}\\% & \mc{1}{c}{R2-31G} \\ 
 \midrule
%STO & 0 &  &  &  &  &  \\
STO-RG & 0.3750236711 & 0 &  &  &  &  \\
STO-R2G & 0.001188713032 & 0.3751362661 & 0 &  &  &  \\
6-31G & 2.050625741 & 1.967532812 & 2.044058927 & 0 &  &  \\
R-31G & 2.218382426 & 1.870166249 & 2.21778158 & 0.1119381976 & 0 &  \\
R2-31G & 1.978898715 & 1.926570559 & 1.980085791 & 0.04202296052 & 0.06880431607  \\ \bottomrule
\end{tabular}%
}
\end{table}

\begin{table}
    \sisetup{round-mode=places,retain-explicit-plus}
    \caption{    \label{tab:r31gl2} $1000 \mathcal{L}_2$. between two core basis functions for carbon in the specified basis set. }
    \centering
    \resizebox{0.48\textwidth}{!}{%
    \begin{tabular}{lS[round-precision=3]S[round-precision=3]S[round-precision=3]S[round-precision=3]S[round-precision=3]S[round-precision=3]}
\toprule
& \mc{1}{c}{STO} & \mc{1}{c}{STO-RG} & \mc{1}{c}{STO-R2G} & \mc{1}{c}{6-31G} & \mc{1}{c}{R-31G} \\% & \mc{1}{c}{R2-31G} \\
\midrule
%STO & 0 &  &  &  &  &  \\
STO-RG & 0.8866113596 & 0 &  &  &  &  \\
STO-R2G & 0.05484085832 & 0.6706186426 & 0 &  &  &  \\
6-31G & 0.05905567611 & 1.02587996 & 0.06165780091 & 0 &  &  \\
R-31G & 0.4666212327 & 0.1902892209 & 0.2743510971 & 0.4480603347 & 0 &  \\
R2-31G & 0.2600827572 & 1.469164044 & 0.4850716219 & 0.3707041183 & 0.9397775772  \\ 
 \bottomrule
\end{tabular}%
}
\end{table}

\begin{table}
    \sisetup{round-mode=places,retain-explicit-plus}
    \caption{    \label{tab:r31gall} Values of key basis function quantities in STO, and errors of other core basis functions compared to the STO basis function. Definitions of column quantities in \Cref{tab:metrics}. Some results are reproduced from \Cref{tab:valueatnuclei} for ease of comparison.}
    \centering
    \begin{tabular}{lS[round-precision=3]S[round-precision=3]S[round-precision=3]}
\toprule
 &  \mc{1}{c}{$\Delta \chi(0)$} & \mc{1}{c}{$ \Delta \frac{\chi^\prime(r)}{\chi(r)}\Big|_{r=0}$} & \mc{1}{c}{$\Delta \mathbb{E}(\chi_1)$}  \\
 \midrule
%STO & 0 &  &  &  &  &  \\
STO & 7.617 & -5.670 & -16.074  \\
$\Delta$STO-6G & -0.460 & +5.670 & +0.006  \\
$\Delta$STO-RG & +0.049352324 & -0.07549779702 &  +0.03160316627  \\
$\Delta$STO-R2G & -0.001446120359 & +0.00304262775 & +0.002069758003\\
$\Delta$6-31G & -0.02825413106 & +5.67 & +0.00767354455  \\
$\Delta$R-31G & +0.1830268102 & -0.1852625119 & +0.02435134162\\
$\Delta$R2-31G & +0.1689146575 & -0.1652512765 & +0.0129145753 \\
\bottomrule
\end{tabular}%

\end{table}

\begin{figure*}
    \centering
    \begin{subfigure}[t]{0.45\textwidth}
        \centering
        \includegraphics[width=\textwidth]{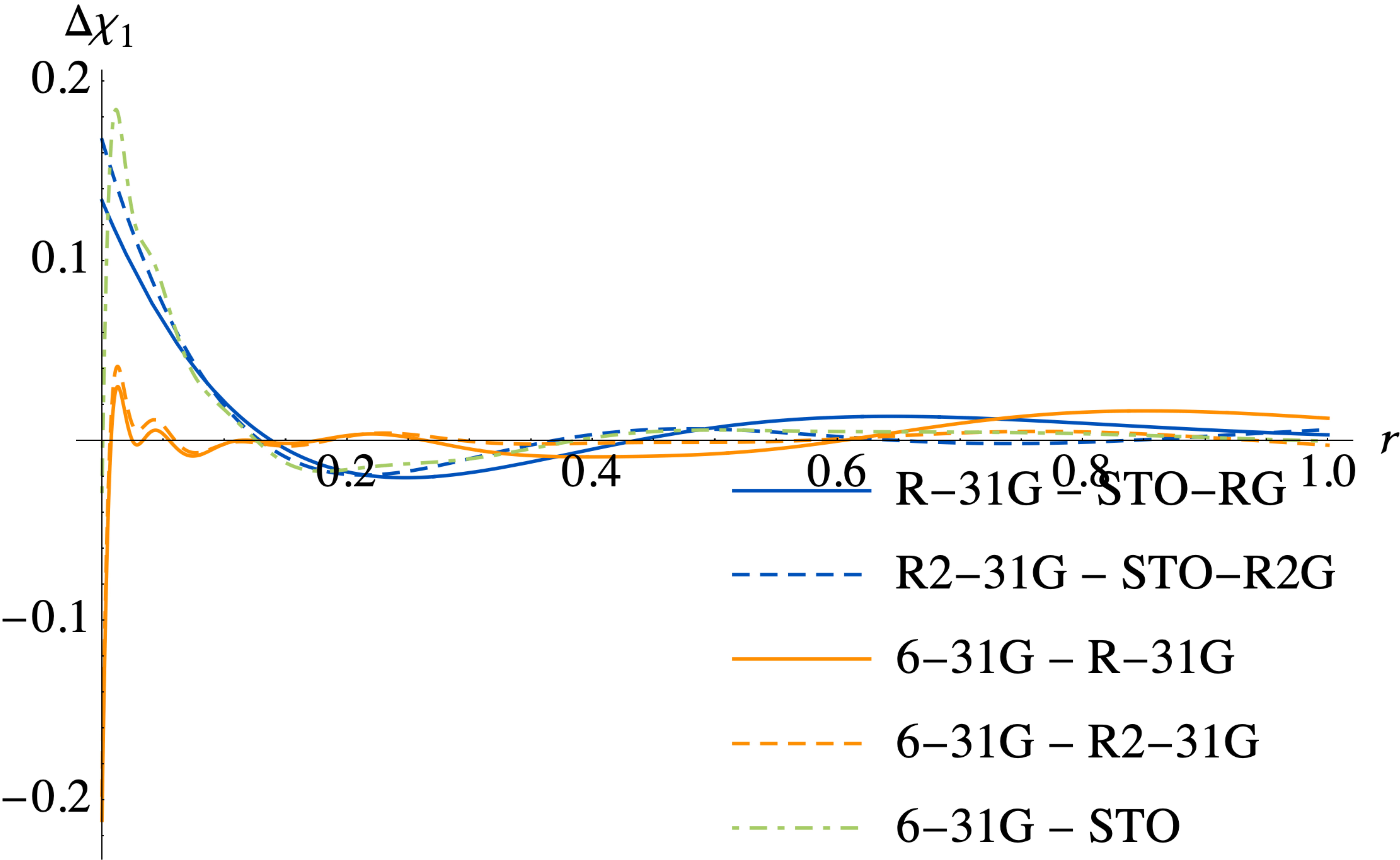}
        \caption{(a) $\Delta \chi_1$, comparisons between STO-R$n$G/6-31G with R$n$-31G}
    \end{subfigure}%
    \hspace{1em}
    \begin{subfigure}[t]{0.45\textwidth}
        \centering
        \includegraphics[width=\textwidth]{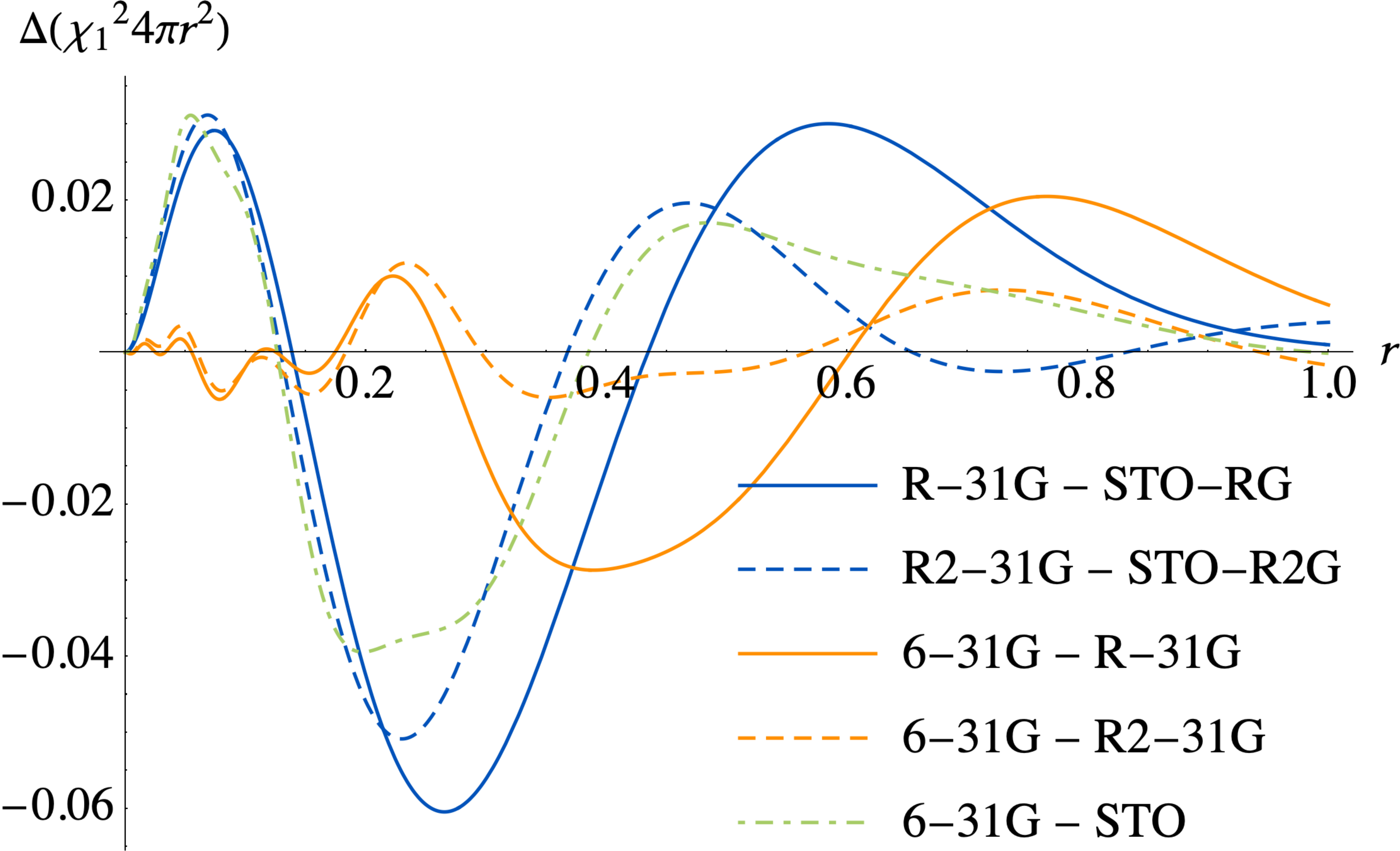}
        \caption{(b) $\Delta 4\pi r^2 (\chi_1^2)$,  comparisons between STO-R$n$G/6-31G with R$n$-31G}
    \end{subfigure}

    \caption{Difference in the basis functions (a) and probability density (b) between the R$n$-31G, STO-R$n$G and 6-31G basis functions for carbon.}
    \label{fig:rn31g}
\end{figure*}

The first mixed ramp-Gaussian basis set was R-31G whose core basis function was derived from 6-31G in the same way that core basis function in STO-RG was derived from the single-zeta STO basis set. The substantial improvement using  STO-R2G compared STO-RG as demonstrated above suggests that an analogous R2-31G basis set (with a core basis function comprised of one ramp and two Gaussian primitives) may be a significant improvement over R-31G. Therefore, we derive the R2-31G basis set for carbon.

In \Cref{tab:rn31gparameters}, the parameters of the R-31G and R2-31G core basis function for carbon are given and contrasted against the STO-RG, STO-R2G and 6-31G parameters. 

When using a ramp with only one Gaussian, \Cref{tab:rn31gparameters} shows the parameters of the fit are quite similar. The parameters of the STO-RG basis set could easily be used as a starting guess for optimising the R-31G basis set parameters.

In contrast, R2-31G and STO-R2G have very different properties. \Cref{tab:rn31gparameters} shows that the second Gaussian has only a very minor contribution to the R2-31G core basis function, unlike for the STO-R2G basis function. \Cref{fig:r231gl1} explores this further by plotting the value of the $\mathcal{L}_1$ fitting parameter as a function of the two Gaussian exponents $\alpha_1$ and $\alpha_2$ (the value of the coefficients is separately optimised). This plot shows that the fit metric is relatively insensitive to the value of $\alpha_2$, supporting only a slight increase in the larger $\alpha_1$ exponent for R2-31G compared to R-31G (4.911 vs 4.545). This behaviour is markedly different from the STO-R2G vs STO-RG fitting parameters (7.827 for STO-R2G vs 4.897 for STO-RG). Therefore, the second Gaussian is much less important when fitting a ramp-Gaussian basis set to a contracted Gaussian basis function rather than Slater function. 

This result raises a major concern: by fitting to 6-31G, R2-31G is probably inheriting the issues of the "6" core basis function and therefore not taking sufficient advantage of the superior performance of the ramp function in the core region. 

To test this, we explore the importance of these fit differences visually and numerically.  \Cref{fig:rn31g} plots the difference between various basis functions and the resulting probability densities. The six basis sets are cross-compared against each other using the $\mathcal{L}_1$ (\Cref{tab:r31gl1}) and $\mathcal{L}_2$ (\Cref{tab:r31gl2}) residuals. Finally, the errors in key basis function properties against STO basis set are shown in \Cref{tab:r31gall}.

\Cref{fig:rn31g} shows that, as expected from the similarity in the R-31G and R2-31G parameters, (6-31G - R-31G) and (6-31G - R2-31G) are similar. Both show very large deviation of R$n$-31G from 6-31G in the very small region around the nucleus with some deviation also at larger distance from the nucleus, with R2-31G performing better in this outer core region. The R$n$-31G and STO-R$n$G basis sets are more different than the R$n$-31G and 6-31G basis sets, with significant deviation particularly at $r<0.15$ Bohr. For the most part, these differences between the R$n$-31G and STO-R$n$G basis sets can be attributed to the differences between their parent 6-31G and STO basis sets. 
%very close to r=0 but small variation . the differences between the R-31G and 6-31G, a and R2-31G basis sets have similar differences from the 6-31G basis set, with   compares the different basis functions to 6-31G, both in terms of the wavefunction and the probability density. Since the R$n$-31G basis functions are close to 6-31G and STO-R$n$G is close to STO, the difference between R$n$-31G and STO-R$n$G mimic the difference between 6-31G and STO for both the wavefunction and the probability density. "R2-31G - STO-R2G" is above "R-31G and STO-RG" for the wavefunction, but below for the probability density. Overall, R$n$-31G is much closer to 6-31G than it is to STO-R$n$G. This is to be expected, due to the differences between 6-31G and STO. 

\Cref{tab:r31gl1} confirms that, as measured by $\mathcal{L}_1$, the nature of the STO-based and 6-31G-based basis sets are slightly different; the value of $1000 \mathcal{L}_1$ between any of (STO, STO-RG, STO-RG) and any of (6-31G, R-31G, R2-31G) is around 2, while within one of the groups, the value is below 0.375. Further, these results quantitatively show that the improvement with the additional Gaussian function is much greater for STO-R$n$G compared to R$n$-31G. 

\Cref{tab:r31gl2} shows that the $\mathcal{L}_2$ metric does not  delineate between the STO-based and 6-31G-based basis sets as clearly as the $\mathcal{L}_1$ metric. This result indicates that the inner core regions of the two groups of basis sets (preferenced by the $\mathcal{L}_1$ metric) are quite different while the outer core regions are much more similar. 

The first numerical column of \Cref{tab:r31gall} shows the difference in $\chi_1(0)$ in different basis sets. Interestingly, the 6-31G basis set is much more similar to the STO basis sets in this metric than the STO-6G basis set is (7.589 for 6-31G, 7.617 for STO, 7.157 for STO-6G); one explanation for this is that the fitting procedure used for 6-31G preferences the energy of the basis function more than the overlap and will therefore tend to have a higher contribution to core basis functions. The R$n$-31G basis sets tend to overestimate the STO value compared to the STO-R$n$G basis sets; this is caused by the slight increase in the contribution of the ramp basis function to the former set of basis sets. 

The second numerical column of \Cref{tab:r31gall} shows the error in the cusp  as calculated by different basis sets. As the 6-31G basis set entirely consists of cuspless Gaussians, it is understandably very poor in this metric. Despite this, the R$n$-31G basis sets fitted to this all-Gaussian core function have very good cusps, only about a factor of two worse from the STO-RG cusp. 

Finally, the third column in \Cref{tab:r31gall} shows the different one-electron energies associated with each core basis function (note that this ignores all contributions from other basis functions). 6-31G has more similar energy to STO than any of the other fitted basis functions, implying that there will be significantly benefits in fully optimising mixed ramp-Gaussian basis sets compared to a simple functional fitting approach. Nevertheless, the R-31G basis set has lower energy than STO-RG, while the R2-31G basis set has higher energy than STO-R2G. These results imply that for moderate accuracy results, rampifying all-Gaussian basis sets by replacing the core basis function with a two-fold contracted (ramp and one Gaussian) basis function may be a useful avenue to pursue.

%%%%%%%%%%%%%%%%%
%%%%%%%%%%%%%%%%%
%%%%%%%%%%%%%%%%%

%%%%%%%%%%%%%%%%%
\subsection{Other all-Gaussian basis sets}
%%%%%%%%%%%%%%%%%
\textcolor{black}{When considering the potential future of mixed ramp-Gaussian basis sets, rampified versions of traditional basis sets is desirable.
Here, we explore the nature of core basis functions within general-purpose and specialised all-Gaussian basis sets, focusing on carbon as a representative example of a first-row (Li-Ne) element. In particular, for each basis set, we want to identify whether the core $1s$ molecular orbital is essentially described by a single basis function (i.e. the basis set is single-core-zeta) or multiple basis sets. A straightforward rampification can only readily be achieved for those basis sets that are clearly single-core-zeta. Though we will not consider it further here, the use of segmented not contracted basis functions is also highly desirable for rampification.} %A single-core-zeta basis set will be significantly easier to rampify. 

Further, we are interested in the similarity of the tightest basis function to the Slater basis function $\mathbb{S}_{5.67}$. This similarity will help determine the suitability of the STO-R$n$G core basis function parameters as a starting point for determining the parameters of the core basis functions in the mixed ramp-Gaussian basis set derived from each all-Gaussian basis set. Given the complexity found above for the R2-31G basis set, full derivation of these new mixed ramp-Gaussian basis sets will be deferred to a later paper to enable an in-depth investigation of the benefits of various rampification strategies for first-row elements. 

\begin{table*}
%\resizebox{0.6\columnwidth}{!}{%
\centering
\footnotesize
    \sisetup{round-mode=places,retain-explicit-plus}
	\caption{\label{tab:CcontrGP} Hartree-Fock calculations on neutral carbon and fit metrics comparing $\mathbb{S}_{5.67}$ to the 1$s$ orbital for different general-purpose and specialised all-Gaussian basis sets are presented. $n_{s}$ represents the number of $s$ contracted basis functions in the basis set and $\epsilon_{1s}$ is the energy for the 1$s$ orbital. $c_{1}$, $c_{2}$ and $c_{3}$ are the coefficients for the three first primitive $s$ functions in each basis set. Other properties are defined in \Cref{tab:metrics}.}  
%	The energy of the $1s$ orbital in neutral triplet carbon, and the contributions of $s$ basis functions to this orbital, where $n_s$ is the number of $s$ contracted basis functions in the molecule, \alert{Describe all the columns properly}. \juan{Juan to complete.}}
	\begin{tabular}{lcS[round-precision=4]S[round-precision=3]S[round-precision=3]S[round-precision=3]S[round-precision=3]S[round-precision=3]S[round-precision=3]S[round-precision=3]S[round-precision=3]S[round-precision=3]S[round-precision=3]S[round-precision=3]S[round-precision=3]S[round-precision=3]S[round-precision=3]S[round-precision=3]S[round-precision=3]S[round-precision=3]S[round-precision=3]S[round-precision=3]S[round-precision=3]}
	\toprule
 	 & & \mc{4}{c}{HF calculation, neutral atom} & \mc{4}{c}{Similarity of $\mathbb{S}_{5.67}$ to $\chi_{1}$} \\
\cmidrule(r){3-6} \cmidrule(r){7-10}
					 & \mc{1}{c}{$n_{s}$} & \mc{1}{c}{$\epsilon_{1s}$}& \mc{1}{c}{$c_1$} & \mc{1}{c}{$c_2$} & \mc{1}{c}{$c_3$} & \mc{1}{c}{1000$\mathcal{L}_1$} & \mc{1}{c}{1000$\mathcal{L}_2$} & \mc{1}{c}{$\Delta \chi_1(r=0)$} & \mc{1}{c}{$\Delta \mathbb{E}(\chi_1)$}  \\
			\midrule
\mc{4}{l}{\bf General-purpose} \\
\mc{2}{l}{\emph{Single-valence-zeta}} \\
pc-0      & 5 & -11.34047 & 0.311145  & 0.743551  & 0.017529  & 19669.7                       & 1350.81  & -7.83909   & -10.9642    \\
pcseg-0   & 5 & -11.34124 & -0.989029 & -0.037070 & 0.012493  & 1.73689  & 2.28836  & 0.422923   & -0.0296558  \\
\vspace{-0.7em} \\
\mc{2}{l}{\emph{Double-valence-zeta}}  \\
6-31G     & 5 & -11.34850 & -0.996612 & -0.019237 & 0.005098  & 2.05063  & 0.590557 & 0.0282541  & -0.00767354 \\
cc-pVDZ   & 3 & -11.35939 & -1.000500 & 0.000360  & 0.003068  & 1.87476  & 1.71157  & -0.0268556 & -0.00944713 \\
pc-1      & 3 & -11.35353 & -0.996411 & -0.021095 & 0.004782  & 2.88831  & 1.1983   & 0.132829   & -0.0153563  \\
pcseg-1   & 3 & -11.35308 & 0.995848  & 0.018812  & -0.004198 & 2.67746  & 1.31188  & 0.145899   & -0.0155056  \\
def2-SVP  & 3 & -11.35709 & 0.992565  & 0.026483  & -0.008975 & 0.888605 & 0.864125 & 0.22985    & -0.0100509  \\
jorge-DZP & 4 & -11.34427 & -0.500506 & -0.585723 & -0.008039 & 6617.79                       & 936.4610 & -4.94205   & -4.67583    \\
\vspace{-0.7em} \\
\mc{2}{l}{\emph{Triple-valence-zeta}}  \\
6-311G    & 4 & -11.35410 & 0.563875  & 0.467121  & 0.001214  & 2449.95                       & 244.7810 & -3.4534    & -1.71559    \\
cc-pVTZ   & 4 & -11.36013 & 0.979089  & 0.000021  & 0.039088  & 1.87716  & 1.75012  & -0.0430308 & -0.00941949 \\
pc-2      & 4 & -11.36140 & 1.000930  & 0.013709  & -0.000033 & 2.05066  & 0.884151 & -0.060891  & -0.00859096 \\
pcseg-2   & 4 & -11.36017 & 0.997926  & 0.006104  & 0.003271  & 2.49521  & 1.39632  & -0.0658438 & -0.0107709  \\
def2-TZVP & 5 & -11.36325 & -0.467990 & -0.613600 & -0.015336 & 8151.31                       & 1050.69  & -5.67049   & -5.5568     \\
jorge-TZP & 5 & -11.34404 & 0.328957  & 0.734416  & 0.026331  & 21050.1                       & 1849.34  & -9.06031   & -12.7584    \\
\vspace{-0.7em} \\
\mc{2}{l}{\emph{Quad-valence-zeta}}  \\
cc-pVQZ   & 5 & -11.36278 & 0.993240  & 0.000023  & 0.015552  & 1.88218  & 1.7408   & -0.118974  & -0.00910842 \\
pc-3      & 6 & -11.36524 & 0.600144  & 0.430895  & 0.011249  & 2039.52                       & 292.8190 & -3.30892   & -1.39571    \\
pcseg-3   & 5 & -11.36587 & -0.509869 & -0.555516 & 0.009609  & 4751.6                        & 603.28   & -4.84057   & -3.12139    \\
def2-QZVP & 7  &   -11.34530        &     0.543312      &    -0.005599       &    0.361414       & 4795.25  & 731.3090 & -4.46985   & -3.45107    \\
jorge-QZP & 6 & -11.34461 & 0.302274  & 0.424850  & -0.365695 & 25434.9                       & 2040.36  & -10.2129   & -15.0017    \\
\vspace{-0.7em} \\

\mc{4}{l}{\bf Specialised} \\
\vspace{-0.7em} \\

\mc{4}{l}{\emph{Core-electron correlation}}  \\

cc-pCVDZ  & 4 & -11.35821 & -1.000775 & 0.000298  & 0.003087  & 1.87476  & 1.71157  & -0.0268556 & -0.00944713 \\
cc-pCVTZ  & 6 & -11.35992 & -0.978987 & -0.000065 & -0.039059 & 1.87716  & 1.75012  & -0.0430308 & -0.00941949 \\
cc-pCVQZ  & 8 & -11.36275 & -0.993429 & -0.000581 & -0.015699 & 1.88218  & 1.7408   & -0.118974  & -0.00910842 \\

\vspace{-0.7em} \\
\mc{4}{l}{\emph{Core-valence correlation }}  \\
cc-pwCVDZ & 4 & -11.35737 & -1.000734 & 0.000239  & 0.003114  & 1.87476  & 1.71157  & -0.0268556 & -0.00944713 \\
cc-pwCVTZ & 6 & -11.35990 & -0.979024 & -0.000016 & -0.039107 & 1.87716  & 1.75012  & -0.0430308 & -0.00941949 \\
cc-pwCVQZ & 8 & -11.36273 & 0.993551  & 0.000573  & 0.015650  & 1.88218  & 1.7408   & -0.118974  & -0.00910842 \\

\vspace{-0.7em} \\
\mc{2}{l}{\emph{Chemical shift}}  \\
pcSseg-0  & 3 & -11.33298 & -0.988972 & -0.037263 & 0.012571  & 1.73689  & 2.28836  & 0.422923   & -0.0296558  \\
pcSseg-1  & 3 & -11.35482 & 0.995867  & 0.018720  & -0.004165 & 2.67746  & 1.31188  & 0.145899   & -0.0155056  \\
pcSseg-2  & 4 & -11.36106 & -0.997932 & -0.006056 & -0.003299 & 2.49521  & 1.39632  & -0.0658438 & -0.0107709  \\
pcSseg-3  & 5 & -11.36654 & 0.509894  & 0.555503  & -0.009638 & 4751.6                        & 603.28   & -4.84057   & -3.12139    \\
%pcSseg-4 &  6 & -11.36524 &  -0.705897 &  -0.314406 &  -0.012874 &  \\

\vspace{-0.7em} \\
\mc{4}{l}{\emph{Spin-spin coupling constants}}  \\
pcJ-0     & 4 & -11.32892 & 0.146579  & 0.456921  & 0.524633  & 1.03588E05   & 3981.01  & -21.0287   & -51.4559    \\
pcJ-1     & 5 & -11.34677 & 0.044168  & 0.612016  & 0.441322  & 5.89623E05   & 6329.91  & -52.3634   & -207.6470   \\
pcJ-2     & 7 & -11.34515 & 0.025494  & 0.314406  & 0.444590  & 1.18837E06 & 7062.52  & -76.3370   & -351.31     \\
pcJ-3     & 9 & -11.34529 & 0.005925  & 0.194280  & 0.267905  & 7.39342E06 & 8571.82  & -198.0840  & -1374.95   \\
%pcJ-4 & 11 & -11.34531 & 0.002408 & 0.166283 & 0.192354 \\
\vspace{-0.7em} \\

\mc{4}{l}{\emph{X-ray parameters}} \\
pcX-1 & 7 & -11.33103 & -0.004504  & -0.033560 & -0.033560 & 1.01118E07 & 8923.27 & -162.2080 & -1884.2   \\
pcX-2 & 10 & -11.343721 & -0.000453 & -0.003507 & -0.018005 & 1.61991E08 & 9702.07 & -669.4780 & -13150.8  \\
pcX-3 & 14 & -11.34545 & 0.000028 & 0.000217 & 0.001139 & 4.61956E09 & 9941.67 & -3607.13  & -128057.0 \\
pcX-4 & 18 & -11.345476 & 0.000003 & 0.000023 & 0.000120 & 6.88573E10 & 9984.7  & -13948.0  & -785108.0 \\
%pcJ-4 & 11 & -11.34531 & 0.002408 & 0.166283 & 0.192354 \\

		\bottomrule
	\end{tabular}
%	}
\end{table*}

The left hand side of \Cref{tab:CcontrGP} presents the contribution of $s$ basis functions to the $1s$ core orbital in neutral triplet carbon atom in a variety of general-purpose and specialised basis sets. The right hand side of \Cref{tab:CcontrGP} compares the $\chi_1$ (tightest) basis function in each basis set against a Slater function with $Z_\textrm{eff}$ $=5.67$. Together, these results can be used to classify each basis set as single-core-zeta or multiple-core-zeta. 

\Cref{tab:CcontrGP} shows that all general-purpose single and double-valence-zeta basis sets are single-core-zeta quality, as illustrated by the large value of $c_1$ compared to $c_2$ and $c_3$ and the small values of $\mathcal{L}_1$, $\mathcal{L}_2$, $\Delta \chi_1(r=0)$ and $\Delta \mathbb{E}(\chi_1)$ for these basis sets. For larger basis sets, there is a mix. The larger representatives from the Dunning basis sets (cc-pVTZ $\&$ cc-pVQZ) are both single-core-zeta quality, as are the Jensen triple-valence-zeta basis sets pc-2 and pcseg-2. However, the triple-valence-zeta def2-TZVP basis set is clearly double-core-zeta, as the larger Jensen basis sets (pc-3 $\&$ pcseg-3). Notably, the Pople style basis set 6-311G is clearly double-core-zeta, despite its name. For these basis sets, the additional flexibility in describing the core $1s$ orbital that is offered by higher core-zeta basis sets will necessarily at least somewhat reduce the flexibility of the basis set in  describing the valence orbitals. As noted by \cite{89GrSc}, in the case of 6-311G, the double-core-zeta nature means that the valence region is not of triple-valence-zeta quality.  %This balance of core and valence electron description is something to be carefully considered when defining new basis sets. 

%\alert{Does having higher core-zeta improve how it models the core region? Then why is pcS-n single core and the other ones multi core. I don't know if we have to explictly say the differences, feel free to ignore me :) } 
\Cref{tab:CcontrGP} shows that the core basis functions in specialised basis sets are more varied. The Dunning core-electron (cc-pCV$n$Z) and core-valance (cc-pwCV$n$Z) correlation families are all single-core-zeta, with very similar values for the coefficients. The pcSseg-$n$ family is single-core-zeta as well, excluding the larger pcSseg-3 which is double-core-zeta. Finally, both the pcJ-$n$ and pcX-$n$ families have a multiple core-zeta representation from the smallest member of the basis set family. The pcJ-$n$ family, for example, starts as triple-core-zeta for pcJ-0, but this increases as the basis set goes up. The high number of core basis functions in the pcJ-n and pcX-n basis sets reflects the importance of the core region to describing spin-spin coupling constants and X-ray spectral parameters respectively; the additional basis functions and the decontracted nature of them means that there is significant flexibility for the core orbitals to adapt to changing valence and nearby electron environments such as that caused by other atoms in molecules.

%All computations were performed by defining both the Slater ($Z_\textrm{eff}$ $=5.67$) and the first $s$ basis function for each basis set, to then calculate the respective values presented in the table. For all these parameters, smaller values are expected as the number of primitive functions in the basis function increases. That can be seen, for example, when comparing $\mathcal{L}_{1}$ and $\mathcal{L}_{2}$ for pc-0 and pc-1 where clearly a better fitting between both functions is obtained with the larger basis set. Nonetheless, this sort of trend seems to be valid for those basis sets that are unambiguously single-core-zeta. The jorge-$n$ZP basis set family has double and triple-core-zeta quality representatives and the residuals become larger as the size of the basis set increases. The same happens with the pcSseg-$n$ family, where the worst values are found for pcSseg-3, a double-core-zeta basis set.

%From the specialised basis sets, both the cc-pCV$n$Z and cc-pwCV$n$Z families show exactly the same values, as the exponents and coefficients for the $1s$ basis function are exactly the same. This implies that an equivalent representation of the core is obtained with both basis sets regardless. The pcJ-$n$ and pcX-$n$ families show both the worst results from the table, due to the fact the core is not singly represented by a basis function, but multiple of them are used instead. 

According to the results in \Cref{tab:CcontrGP}, the similarity between the core basis function in 6-31G and $\mathbb{S}_{5.67}$ is about the same as the similarity between any single-core-zeta all-Gaussian core basis function and the $\mathbb{S}_{5.67}$. Therefore, it is likely that if a single Gaussian and single ramp are used, the parameters of a core mixed ramp-Gaussian basis functions derived from single-core-zeta all-Gaussian basis sets are likely to be similar to those in R-31G and STO-RG.

\section{Conclusions and Implications}
Historically, core $1s$ orbitals in first-row atoms Li to Ne have been described by either Slater functions, which more accurately represent the true wavefunction, or Gaussian functions, which are faster computationally. We have recently introduced a new type of core basis function, the mixed ramp-Gaussian basis function, which has a better representation of the near-nuclei region of the wavefunction than Gaussian functions with integrals that can be computed in times similar to all-Gaussian basis functions. 

The central motivation for mixed ramp-Gaussian basis sets has crystallised recently to a focus on core-electron dependent properties. 
While initially we hoped that mixed ramp-Gaussian basis sets may be useful for faster calculations of big biomolecular systems (e.g. proteins) \cite{14McGiGi}, calculations from McKemmish (2015) \cite{15Mc} make it clear that the potential time savings from even removing the core basis functions entirely are still modest (on order of 25\%). On the other hand, improvements in the description of the core electron region, demonstrated in McKemmish and Gilbert (2015) \cite{15McGi} and this paper, through the use of mixed ramp-Gaussian basis sets over all-Gaussian basis sets are radical and have the potential to revolutionise the computational prediction of core electron properties such as NMR spectroscopy parameters. 

In this paper, we start development of new mixed ramp-Gaussian basis sets by modelling Slater functions with a core ramp-Gaussian basis function (with 1 ramp and $n$ Gaussians) yielding the STO-R$n$G basis sets analogous to the early STO-$n$G basis sets. Metric 1 - minimising $\mathcal{L}_1 = \Braket{\mathcal{S}^2-\mathcal{RG}^2|\mathcal{S}^2-\mathcal{RG}^2}$ - was selected for the fitting, as it prioritises the core region, the region of interest.  

A brief investigation into the rampfication of all-Gaussian basis sets was also conducted, by comparing the existing R-31G basis set to the novel STO-R$n$G and R2-31G basis functions. Unlike for the Slater case, the additional Gaussian in R2-31G did not significantly improve the fit quality. 

%It was found that only one Gaussian primitive was needed to accurately capture the core of the 6-31G basis function, and that if an additional Gaussian was added, it contributed to the description of the valence region. These findings will influence the development of more mixed ramp-Gaussian basis sets.

Moving forward, we aim to start construction of new ramp-Gaussian basis sets by deriving basis sets from a variety of parent all-Gaussian basis sets. The best methodology for this future rampification of the core basis function is unclear; should a single ramp and Gaussian be fitted against the existing all-Gaussian core basis function, or should a more complicated procedure be used to fit a ramp plus two Gaussians to a more accurate core basis function (such as a Slater type orbital)?  % using a similar fitting method as developed in this paper.  % The first mixed ramp-Gaussian basis set family, R-31G (based on 6-31G), used only one ramp and one Gaussian in its core basis function. Our results here showed no major advantage to adding an additional Gaussian to form R2-31G when fitting a ramp-Gaussian basis funciton against an all-Gaussian core basis function. However, the results 

%However, given the results in this paper and the now key motivating purpose for mixed ramp-Gaussian basis sets, moving forward we will use more Gaussians, usually replacing the core Gaussian basis function with one ramp and two Gaussian functions. \alert{Does this need to be changed in light of R2-31G?}

In any case, straightforward rampification of mixed ramp-Gaussian basis sets is only possible if a single core basis function can be clearly identified in a given all-Gaussian basis set. This paper examined a variety of general-purpose and specialised basis sets for carbon to identify single-core-zeta and multiple-core-zeta basis sets by looking at relative importance of the three tightest $s$ basis functions to the $1s$ core orbital in an atomic calculation as well as the similarity between the tightest basis function and the minimal Slater basis function for this orbital. We find that: 
\begin{itemize}
\item 6-31G, the Dunning basis sets (including cc-pV$n$Z, cc-pCV$n$Z and cc-pwCV$n$Z), double and triple-zeta Jensen basis sets (including pc-$n$, pcseg-$n$ and pcS-$n$) and def2-SVP have only one major core basis function (i.e. can be classified as single-core-zeta) and are thus seem to be good candidates for straightforward rampification in the future;
    \item 6-311G, Jensen basis sets of quadrupole zeta quality, def2 basis sets of triple and quadrupole zeta quality and jorge basis sets generally have large contributsion of two basis functions to the $1s$ orbital and will be more complicated to rampify;
    \item the Jensen pcJ-$n$ and pcX-$n$ basis sets specialised for spin-spin coupling constants and X-ray properties respectively have such significant decontraction in the core that direct rampification is impractical and a full re-optimisation of the basis set should be undertaken. 
\end{itemize}%\alert{Juan: add description of results here}. 

The results for 6-311G are particularly notable. The basis set name implies it is triple-valence-zeta. Despite this, as previously observed \cite{89GrSc} but insufficiently recognised, it is double-core-zeta and thus not of true triple-valence-zeta quality. 

Future work will utilise the results in this paper to rampify the identified single-core-zeta all-Gaussian basis sets to produce mixed ramp-Gaussian basis sets that can more accurately model core electron properties such as X-ray and NMR spectral parameters. \textcolor{black}{Additionally, the scope of the mixed ramp-Gaussian basis sets derived from all-Gaussian basis sets will be expanded to include second-row elements. This involves the rampification of the core 2$s$ and 2$p$ basis functions.}
%The parameters of the STO-RG basis set can serve as the initial guess for the parameters of the smaller rampified mixed ramp-Gaussian basis sets developed from the single-core-zeta all-Gaussian basis sets identified in this paper, while the parameters of the STO-R2G basis set can be used 

%We will discuss importance of segmented vs contracted basis sets in next paper I think.

\section*{Acknowledgements} This research was undertaken with the assistance of resources from the National Computational Infrastructure (NCI Australia), an NCRIS enabled capability supported by the Australian Government.

The authors declare no conflicts of interest.

%\printbibliography[heading=bibintoc]
\bibliography{Paper_Slaters}

\providecommand{\latin}[1]{#1}
\makeatletter
\providecommand{\doi}
  {\begingroup\let\do\@makeother\dospecials
  \catcode`\{=1 \catcode`\}=2 \doi@aux}
\providecommand{\doi@aux}[1]{\endgroup\texttt{#1}}
\makeatother
\providecommand*\mcitethebibliography{\thebibliography}
\csname @ifundefined\endcsname{endmcitethebibliography}
  {\let\endmcitethebibliography\endthebibliography}{}
\begin{mcitethebibliography}{56}
\providecommand*\natexlab[1]{#1}
\providecommand*\mciteSetBstSublistMode[1]{}
\providecommand*\mciteSetBstMaxWidthForm[2]{}
\providecommand*\mciteBstWouldAddEndPuncttrue
  {\def\EndOfBibitem{\unskip.}}
\providecommand*\mciteBstWouldAddEndPunctfalse
  {\let\EndOfBibitem\relax}
\providecommand*\mciteSetBstMidEndSepPunct[3]{}
\providecommand*\mciteSetBstSublistLabelBeginEnd[3]{}
\providecommand*\EndOfBibitem{}
\mciteSetBstSublistMode{f}
\mciteSetBstMaxWidthForm{subitem}{(\alph{mcitesubitemcount})}
\mciteSetBstSublistLabelBeginEnd
  {\mcitemaxwidthsubitemform\space}
  {\relax}
  {\relax}

\bibitem[Br{\'e}mond \latin{et~al.}(2016)Br{\'e}mond, Savarese, Su,
  P{\'e}rez-Jim{\'e}nez, Xu, Sancho-Garc{\'\i}a, and Adamo]{16BrSaSu}
Br{\'e}mond,~{\'E}.; Savarese,~M.; Su,~N.~Q.; P{\'e}rez-Jim{\'e}nez,~{\'A}.~J.;
  Xu,~X.; Sancho-Garc{\'\i}a,~J.~C.; Adamo,~C. \emph{Journal of chemical theory
  and computation} \textbf{2016}, \emph{12}, 459--465\relax
\mciteBstWouldAddEndPuncttrue
\mciteSetBstMidEndSepPunct{\mcitedefaultmidpunct}
{\mcitedefaultendpunct}{\mcitedefaultseppunct}\relax
\EndOfBibitem
\bibitem[Karton(2016)]{thermochem}
Karton,~A. \emph{Wiley Interdisciplinary Reviews: Computational Molecular
  Science} \textbf{2016}, \emph{6}, 292--310\relax
\mciteBstWouldAddEndPuncttrue
\mciteSetBstMidEndSepPunct{\mcitedefaultmidpunct}
{\mcitedefaultendpunct}{\mcitedefaultseppunct}\relax
\EndOfBibitem
\bibitem[Peterson \latin{et~al.}(2012)Peterson, Feller, and Dixon]{12PeFeDi}
Peterson,~K.~A.; Feller,~D.; Dixon,~D.~A. \emph{Theoretical Chemistry Accounts}
  \textbf{2012}, \emph{131}, 1079\relax
\mciteBstWouldAddEndPuncttrue
\mciteSetBstMidEndSepPunct{\mcitedefaultmidpunct}
{\mcitedefaultendpunct}{\mcitedefaultseppunct}\relax
\EndOfBibitem
\bibitem[Cheng \latin{et~al.}(2015)Cheng, Zhang, Chung, Xu, and Wu]{reactivity}
Cheng,~G.~J.; Zhang,~X.; Chung,~L.~W.; Xu,~L.; Wu,~Y.~D. \emph{Journal of the
  American Chemical Society} \textbf{2015}, \emph{137}, 1706--1725\relax
\mciteBstWouldAddEndPuncttrue
\mciteSetBstMidEndSepPunct{\mcitedefaultmidpunct}
{\mcitedefaultendpunct}{\mcitedefaultseppunct}\relax
\EndOfBibitem
\bibitem[Pople(1998)]{pople_lecture}
Pople,~J.~A. {Quantum Chemical Models}. 1998\relax
\mciteBstWouldAddEndPuncttrue
\mciteSetBstMidEndSepPunct{\mcitedefaultmidpunct}
{\mcitedefaultendpunct}{\mcitedefaultseppunct}\relax
\EndOfBibitem
\bibitem[Goerigk and Mehta(2019)Goerigk, and Mehta]{DFT_zoo}
Goerigk,~L.; Mehta,~N. \emph{Australian Journal of Chemistry} \textbf{2019},
  \relax
\mciteBstWouldAddEndPunctfalse
\mciteSetBstMidEndSepPunct{\mcitedefaultmidpunct}
{}{\mcitedefaultseppunct}\relax
\EndOfBibitem
\bibitem[Gersten and Frederick(2001)Gersten, and Frederick]{valence_electrons}
Gersten,~J.~I.; Frederick,~S.~W. \emph{{The Physics and Chemistry of
  Materials}}, 1st ed.; Wiley-Interscience, 2001; p 856\relax
\mciteBstWouldAddEndPuncttrue
\mciteSetBstMidEndSepPunct{\mcitedefaultmidpunct}
{\mcitedefaultendpunct}{\mcitedefaultseppunct}\relax
\EndOfBibitem
\bibitem[Darbeau(2006)]{nmr_review}
Darbeau,~R. \emph{Applied Spectroscopy Reviews} \textbf{2006}, \emph{41},
  401--425\relax
\mciteBstWouldAddEndPuncttrue
\mciteSetBstMidEndSepPunct{\mcitedefaultmidpunct}
{\mcitedefaultendpunct}{\mcitedefaultseppunct}\relax
\EndOfBibitem
\bibitem[Jensen(2008)]{pcS}
Jensen,~F. \emph{Journal of Chemical Theory and Computation} \textbf{2008},
  \emph{4}, 719--727, PMID: 26621087\relax
\mciteBstWouldAddEndPuncttrue
\mciteSetBstMidEndSepPunct{\mcitedefaultmidpunct}
{\mcitedefaultendpunct}{\mcitedefaultseppunct}\relax
\EndOfBibitem
\bibitem[Jensen(2015)]{pcSseg-n}
Jensen,~F. \emph{Journal of Chemical Theory and Computation} \textbf{2015},
  \emph{11}, 132--138\relax
\mciteBstWouldAddEndPuncttrue
\mciteSetBstMidEndSepPunct{\mcitedefaultmidpunct}
{\mcitedefaultendpunct}{\mcitedefaultseppunct}\relax
\EndOfBibitem
\bibitem[Jensen(2010)]{pcJ}
Jensen,~F. \emph{Theor. Chem. Acc.} \textbf{2010}, \emph{126}, 371--382\relax
\mciteBstWouldAddEndPuncttrue
\mciteSetBstMidEndSepPunct{\mcitedefaultmidpunct}
{\mcitedefaultendpunct}{\mcitedefaultseppunct}\relax
\EndOfBibitem
\bibitem[Roessler and Salvadori(2018)Roessler, and Salvadori]{epr_review}
Roessler,~M.~M.; Salvadori,~E. \emph{Chemical Society Reviews} \textbf{2018},
  \emph{47}, 2534--2553\relax
\mciteBstWouldAddEndPuncttrue
\mciteSetBstMidEndSepPunct{\mcitedefaultmidpunct}
{\mcitedefaultendpunct}{\mcitedefaultseppunct}\relax
\EndOfBibitem
\bibitem[Felton(2003)]{auger_review}
Felton,~M.~J. \emph{Analytical Chemistry} \textbf{2003}, \emph{75}, 269 A--271
  A\relax
\mciteBstWouldAddEndPuncttrue
\mciteSetBstMidEndSepPunct{\mcitedefaultmidpunct}
{\mcitedefaultendpunct}{\mcitedefaultseppunct}\relax
\EndOfBibitem
\bibitem[Gibb(1976)]{gibb_mossbook}
Gibb,~T.~C. \emph{{Principles of M{\"{o}}ssbauer Spectroscopy}}, 1st ed.;
  Springer: Norwich, 1976; p 254\relax
\mciteBstWouldAddEndPuncttrue
\mciteSetBstMidEndSepPunct{\mcitedefaultmidpunct}
{\mcitedefaultendpunct}{\mcitedefaultseppunct}\relax
\EndOfBibitem
\bibitem[Loh and Leone(2013)Loh, and Leone]{xray_review1}
Loh,~Z.~H.; Leone,~S.~R. \emph{Journal of Physical Chemistry Letters}
  \textbf{2013}, \emph{4}, 292--302\relax
\mciteBstWouldAddEndPuncttrue
\mciteSetBstMidEndSepPunct{\mcitedefaultmidpunct}
{\mcitedefaultendpunct}{\mcitedefaultseppunct}\relax
\EndOfBibitem
\bibitem[Kraus \latin{et~al.}(2018)Kraus, Z{\"{u}}rch, Cushing, Neumark, and
  Leone]{xray_review2}
Kraus,~P.~M.; Z{\"{u}}rch,~M.; Cushing,~S.~K.; Neumark,~D.~M.; Leone,~S.~R.
  \emph{Nature Reviews Chemistry} \textbf{2018}, \emph{2}, 82--94\relax
\mciteBstWouldAddEndPuncttrue
\mciteSetBstMidEndSepPunct{\mcitedefaultmidpunct}
{\mcitedefaultendpunct}{\mcitedefaultseppunct}\relax
\EndOfBibitem
\bibitem[Ambroise and Jensen(2018)Ambroise, and Jensen]{pcX}
Ambroise,~M.~A.; Jensen,~F. \emph{Journal of Chemical Theory and Computation}
  \textbf{2018}, \emph{15}, 325--337\relax
\mciteBstWouldAddEndPuncttrue
\mciteSetBstMidEndSepPunct{\mcitedefaultmidpunct}
{\mcitedefaultendpunct}{\mcitedefaultseppunct}\relax
\EndOfBibitem
\bibitem[Jensen(2007)]{jensen_book}
Jensen,~F. \emph{Annual Reports on the Progress of Chemistry - Section C}, 2nd
  ed.; Jhon Wiley {\&} Sons: Chichester, 2007; Vol.~90; p 583\relax
\mciteBstWouldAddEndPuncttrue
\mciteSetBstMidEndSepPunct{\mcitedefaultmidpunct}
{\mcitedefaultendpunct}{\mcitedefaultseppunct}\relax
\EndOfBibitem
\bibitem[Slater(1930)]{slater1930atomic}
Slater,~J.~C. \emph{Physical Review} \textbf{1930}, \emph{36}, 57\relax
\mciteBstWouldAddEndPuncttrue
\mciteSetBstMidEndSepPunct{\mcitedefaultmidpunct}
{\mcitedefaultendpunct}{\mcitedefaultseppunct}\relax
\EndOfBibitem
\bibitem[Hoggan \latin{et~al.}(2011)Hoggan, Belen~Ru{\'{i}}z, and
  Ozdogan]{sto_comp_review}
Hoggan,~P.~E.; Belen~Ru{\'{i}}z,~M.; Ozdogan,~T. \emph{Quantum Frontiers of
  Atoms and Molecules (Chemistry Research and Applications)}; Nova Publishing
  Inc.: New York, 2011; Chapter 4, p 673\relax
\mciteBstWouldAddEndPuncttrue
\mciteSetBstMidEndSepPunct{\mcitedefaultmidpunct}
{\mcitedefaultendpunct}{\mcitedefaultseppunct}\relax
\EndOfBibitem
\bibitem[Harris and Michels(1965)Harris, and Michels]{sto_comp1}
Harris,~F.~E.; Michels,~H.~H. \emph{The Journal of Chemical Physics}
  \textbf{1965}, \emph{43}\relax
\mciteBstWouldAddEndPuncttrue
\mciteSetBstMidEndSepPunct{\mcitedefaultmidpunct}
{\mcitedefaultendpunct}{\mcitedefaultseppunct}\relax
\EndOfBibitem
\bibitem[Harris and Michels(1966)Harris, and Michels]{sto_comp2}
Harris,~F.~E.; Michels,~H.~H. \emph{The Journal of Chemical Physics}
  \textbf{1966}, \emph{45}, 116--123\relax
\mciteBstWouldAddEndPuncttrue
\mciteSetBstMidEndSepPunct{\mcitedefaultmidpunct}
{\mcitedefaultendpunct}{\mcitedefaultseppunct}\relax
\EndOfBibitem
\bibitem[Talman(1984)]{sto_comp3}
Talman,~J.~D. \emph{The Journal of Chemical Physics} \textbf{1984}, \emph{80},
  2000--2008\relax
\mciteBstWouldAddEndPuncttrue
\mciteSetBstMidEndSepPunct{\mcitedefaultmidpunct}
{\mcitedefaultendpunct}{\mcitedefaultseppunct}\relax
\EndOfBibitem
\bibitem[Filter and Steinborn(1978)Filter, and Steinborn]{sto_comp4}
Filter,~E.; Steinborn,~E.~O. \emph{Physical Review A} \textbf{1978}, \emph{18},
  11\relax
\mciteBstWouldAddEndPuncttrue
\mciteSetBstMidEndSepPunct{\mcitedefaultmidpunct}
{\mcitedefaultendpunct}{\mcitedefaultseppunct}\relax
\EndOfBibitem
\bibitem[Huzinaga(1985)]{sto_inefficient}
Huzinaga,~S. \emph{Computer Physics Reports} \textbf{1985}, \emph{2},
  281--339\relax
\mciteBstWouldAddEndPuncttrue
\mciteSetBstMidEndSepPunct{\mcitedefaultmidpunct}
{\mcitedefaultendpunct}{\mcitedefaultseppunct}\relax
\EndOfBibitem
\bibitem[te~Velde \latin{et~al.}(2001)te~Velde, Bickelhaupt, Baerends,
  Fonseca~Guerra, van Gisbergen, Snijders, and Ziegler]{adf}
te~Velde,~G.; Bickelhaupt,~F.~M.; Baerends,~E.~J.; Fonseca~Guerra,~C.; van
  Gisbergen,~S.~J.; Snijders,~J.~G.; Ziegler,~T. \emph{Journal of Computational
  Chemistry} \textbf{2001}, \emph{22}, 931--967\relax
\mciteBstWouldAddEndPuncttrue
\mciteSetBstMidEndSepPunct{\mcitedefaultmidpunct}
{\mcitedefaultendpunct}{\mcitedefaultseppunct}\relax
\EndOfBibitem
\bibitem[Boys(1950)]{Boys50}
Boys,~S.~F. \emph{Proc. Roy. Soc. (London)} \textbf{1950}, \emph{A200},
  542--554\relax
\mciteBstWouldAddEndPuncttrue
\mciteSetBstMidEndSepPunct{\mcitedefaultmidpunct}
{\mcitedefaultendpunct}{\mcitedefaultseppunct}\relax
\EndOfBibitem
\bibitem[Jalbout \latin{et~al.}(2004)Jalbout, Nazari, and Turker]{gaussians}
Jalbout,~A.~F.; Nazari,~F.; Turker,~L. \emph{Journal of Molecular Structure:
  Theochem} \textbf{2004}, \emph{671}, 1--21\relax
\mciteBstWouldAddEndPuncttrue
\mciteSetBstMidEndSepPunct{\mcitedefaultmidpunct}
{\mcitedefaultendpunct}{\mcitedefaultseppunct}\relax
\EndOfBibitem
\bibitem[Gill \latin{et~al.}(1994)Gill, Johnson, and Pople]{gill1994simple}
Gill,~P.~M.; Johnson,~B.~G.; Pople,~J.~A. \emph{Chem. Phys. Lett.}
  \textbf{1994}, \emph{217}, 65--68\relax
\mciteBstWouldAddEndPuncttrue
\mciteSetBstMidEndSepPunct{\mcitedefaultmidpunct}
{\mcitedefaultendpunct}{\mcitedefaultseppunct}\relax
\EndOfBibitem
\bibitem[Hehre \latin{et~al.}(1969)Hehre, Stewart, and Pople]{sto-nG}
Hehre,~W.~J.; Stewart,~R.~F.; Pople,~J.~A. \emph{The Journal of Chemical
  Physics} \textbf{1969}, \emph{51}, 2657--2664\relax
\mciteBstWouldAddEndPuncttrue
\mciteSetBstMidEndSepPunct{\mcitedefaultmidpunct}
{\mcitedefaultendpunct}{\mcitedefaultseppunct}\relax
\EndOfBibitem
\bibitem[Dobbs and Hehre(1987)Dobbs, and Hehre]{3-21G}
Dobbs,~K.~D.; Hehre,~W.~J. \emph{Journal of Computational Chemistry}
  \textbf{1987}, \emph{8}, 861--879\relax
\mciteBstWouldAddEndPuncttrue
\mciteSetBstMidEndSepPunct{\mcitedefaultmidpunct}
{\mcitedefaultendpunct}{\mcitedefaultseppunct}\relax
\EndOfBibitem
\bibitem[Hehre \latin{et~al.}(1972)Hehre, Ditchfield, and Pople]{6-31G}
Hehre,~W.~J.; Ditchfield,~K.; Pople,~J.~A. \emph{The Journal of Chemical
  Physics} \textbf{1972}, \emph{56}, 2257--2261\relax
\mciteBstWouldAddEndPuncttrue
\mciteSetBstMidEndSepPunct{\mcitedefaultmidpunct}
{\mcitedefaultendpunct}{\mcitedefaultseppunct}\relax
\EndOfBibitem
\bibitem[Rassolov \latin{et~al.}(1998)Rassolov, Pople, Ratner, and
  Windus]{6-31Gstar}
Rassolov,~V.~A.; Pople,~J.~A.; Ratner,~M.~A.; Windus,~T.~L. \emph{Journal of
  Chemical Physics} \textbf{1998}, \emph{109}, 1223--1229\relax
\mciteBstWouldAddEndPuncttrue
\mciteSetBstMidEndSepPunct{\mcitedefaultmidpunct}
{\mcitedefaultendpunct}{\mcitedefaultseppunct}\relax
\EndOfBibitem
\bibitem[Krishnan \latin{et~al.}(1980)Krishnan, Binkley, Seeger, and
  Pople]{6-311G_star}
Krishnan,~R.; Binkley,~J.~S.; Seeger,~R.; Pople,~J.~A. \emph{The Journal of
  Chemical Physics} \textbf{1980}, \emph{72}, 650--654\relax
\mciteBstWouldAddEndPuncttrue
\mciteSetBstMidEndSepPunct{\mcitedefaultmidpunct}
{\mcitedefaultendpunct}{\mcitedefaultseppunct}\relax
\EndOfBibitem
\bibitem[Frisch \latin{et~al.}(1984)Frisch, Pople, and Binkley]{6-31Gd}
Frisch,~M.~J.; Pople,~J.~A.; Binkley,~J.~S. \emph{The Journal of Chemical
  Physics} \textbf{1984}, \emph{80}, 3265--3269\relax
\mciteBstWouldAddEndPuncttrue
\mciteSetBstMidEndSepPunct{\mcitedefaultmidpunct}
{\mcitedefaultendpunct}{\mcitedefaultseppunct}\relax
\EndOfBibitem
\bibitem[Dunning(1989)]{cc-pVnZ_B-Ne}
Dunning,~T.~H. \emph{The Journal of Chemical Physics} \textbf{1989}, \emph{90},
  1007--1023\relax
\mciteBstWouldAddEndPuncttrue
\mciteSetBstMidEndSepPunct{\mcitedefaultmidpunct}
{\mcitedefaultendpunct}{\mcitedefaultseppunct}\relax
\EndOfBibitem
\bibitem[Woon and Dunning(1995)Woon, and Dunning]{cc-pCVnZ}
Woon,~D.~E.; Dunning,~T.~H. \emph{The Journal of Chemical Physics}
  \textbf{1995}, \emph{103}, 4572--4585\relax
\mciteBstWouldAddEndPuncttrue
\mciteSetBstMidEndSepPunct{\mcitedefaultmidpunct}
{\mcitedefaultendpunct}{\mcitedefaultseppunct}\relax
\EndOfBibitem
\bibitem[Peterson and Dunning(2002)Peterson, and Dunning]{cc-pwCVTZ}
Peterson,~K.~A.; Dunning,~T.~H. \emph{Journal of Chemical Physics}
  \textbf{2002}, \emph{117}, 10548--10560\relax
\mciteBstWouldAddEndPuncttrue
\mciteSetBstMidEndSepPunct{\mcitedefaultmidpunct}
{\mcitedefaultendpunct}{\mcitedefaultseppunct}\relax
\EndOfBibitem
\bibitem[Jensen(2001)]{pc-n}
Jensen,~F. \emph{Journal of Chemical Physics} \textbf{2001}, \emph{115},
  9113--9125\relax
\mciteBstWouldAddEndPuncttrue
\mciteSetBstMidEndSepPunct{\mcitedefaultmidpunct}
{\mcitedefaultendpunct}{\mcitedefaultseppunct}\relax
\EndOfBibitem
\bibitem[Jensen(2014)]{pcseg-n}
Jensen,~F. \emph{Journal of Chemical Theory and Computation} \textbf{2014},
  \emph{10}, 1074--1085\relax
\mciteBstWouldAddEndPuncttrue
\mciteSetBstMidEndSepPunct{\mcitedefaultmidpunct}
{\mcitedefaultendpunct}{\mcitedefaultseppunct}\relax
\EndOfBibitem
\bibitem[Jensen(2013)]{atomic_orbitals}
Jensen,~F. \emph{Wiley Interdisciplinary Reviews: Computational Molecular
  Science} \textbf{2013}, \emph{3}, 273--295\relax
\mciteBstWouldAddEndPuncttrue
\mciteSetBstMidEndSepPunct{\mcitedefaultmidpunct}
{\mcitedefaultendpunct}{\mcitedefaultseppunct}\relax
\EndOfBibitem
\bibitem[Nagy and Jensen(2017)Nagy, and Jensen]{Nagy2017BasisChemistry}
Nagy,~B.; Jensen,~F. \emph{Reviews in Computational Chemistry, Volume 30}
  \textbf{2017}, 93--149\relax
\mciteBstWouldAddEndPuncttrue
\mciteSetBstMidEndSepPunct{\mcitedefaultmidpunct}
{\mcitedefaultendpunct}{\mcitedefaultseppunct}\relax
\EndOfBibitem
\bibitem[Feller(1996)]{BSE1}
Feller,~D. \emph{Journal of Computational Chemistry} \textbf{1996}, \emph{17},
  1571--1586\relax
\mciteBstWouldAddEndPuncttrue
\mciteSetBstMidEndSepPunct{\mcitedefaultmidpunct}
{\mcitedefaultendpunct}{\mcitedefaultseppunct}\relax
\EndOfBibitem
\bibitem[Schuchardt \latin{et~al.}(2007)Schuchardt, Didier, Elsethagen, Sun,
  Gurumoorthi, Chase, Li, and Windus]{BSE2}
Schuchardt,~K.~L.; Didier,~B.~T.; Elsethagen,~T.; Sun,~L.; Gurumoorthi,~V.;
  Chase,~J.; Li,~J.; Windus,~T.~L. \emph{Journal of Chemical Information and
  Modeling} \textbf{2007}, \emph{47}, 1045--1052, PMID: 17428029\relax
\mciteBstWouldAddEndPuncttrue
\mciteSetBstMidEndSepPunct{\mcitedefaultmidpunct}
{\mcitedefaultendpunct}{\mcitedefaultseppunct}\relax
\EndOfBibitem
\bibitem[McKemmish and Gill(2012)McKemmish, and Gill]{12McGi}
McKemmish,~L.~K.; Gill,~P.~M. \emph{Journal of Chemical Theory and Computation}
  \textbf{2012}, \emph{8}, 4891--4898\relax
\mciteBstWouldAddEndPuncttrue
\mciteSetBstMidEndSepPunct{\mcitedefaultmidpunct}
{\mcitedefaultendpunct}{\mcitedefaultseppunct}\relax
\EndOfBibitem
\bibitem[Wachters(1970)]{wachters_basis}
Wachters,~A. J.~H. \emph{Journal of Chemical Physics} \textbf{1970}, \emph{52},
  1033--1036\relax
\mciteBstWouldAddEndPuncttrue
\mciteSetBstMidEndSepPunct{\mcitedefaultmidpunct}
{\mcitedefaultendpunct}{\mcitedefaultseppunct}\relax
\EndOfBibitem
\bibitem[Neese(2002)]{cp_ppp}
Neese,~F. \emph{Inorganica Chimica Acta} \textbf{2002}, \emph{337},
  181--192\relax
\mciteBstWouldAddEndPuncttrue
\mciteSetBstMidEndSepPunct{\mcitedefaultmidpunct}
{\mcitedefaultendpunct}{\mcitedefaultseppunct}\relax
\EndOfBibitem
\bibitem[Laasner \latin{et~al.}(2018)Laasner, Huhn, Colell, Theis, Yu, Warren,
  and Blum]{num_ato}
Laasner,~R.; Huhn,~W.; Colell,~J.; Theis,~T.; Yu,~V.; Warren,~W.; Blum,~V.
  \textbf{2018}, \relax
\mciteBstWouldAddEndPunctfalse
\mciteSetBstMidEndSepPunct{\mcitedefaultmidpunct}
{}{\mcitedefaultseppunct}\relax
\EndOfBibitem
\bibitem[Zhang \latin{et~al.}(2013)Zhang, Ren, Rinke, Blum, and
  Scheffler]{13ZhReRi.NAO}
Zhang,~I.~Y.; Ren,~X.; Rinke,~P.; Blum,~V.; Scheffler,~M. \emph{New Journal of
  Physics} \textbf{2013}, \emph{15}, 123033\relax
\mciteBstWouldAddEndPuncttrue
\mciteSetBstMidEndSepPunct{\mcitedefaultmidpunct}
{\mcitedefaultendpunct}{\mcitedefaultseppunct}\relax
\EndOfBibitem
\bibitem[McKemmish \latin{et~al.}(2014)McKemmish, Gilbert, and Gill]{14McGiGi}
McKemmish,~L.~K.; Gilbert,~A.~T.; Gill,~P.~M. \emph{Journal of Chemical Theory
  and Computation} \textbf{2014}, \emph{10}, 4369--4376\relax
\mciteBstWouldAddEndPuncttrue
\mciteSetBstMidEndSepPunct{\mcitedefaultmidpunct}
{\mcitedefaultendpunct}{\mcitedefaultseppunct}\relax
\EndOfBibitem
\bibitem[Bishop(1964)]{ramps_1}
Bishop,~D.~M. \emph{The Journal of Chemical Physics} \textbf{1964}, \emph{40},
  1322--1325\relax
\mciteBstWouldAddEndPuncttrue
\mciteSetBstMidEndSepPunct{\mcitedefaultmidpunct}
{\mcitedefaultendpunct}{\mcitedefaultseppunct}\relax
\EndOfBibitem
\bibitem[Bishop(1968)]{ramps_2}
Bishop,~D.~M. \emph{The Journal of Chemical Physics} \textbf{1968}, \emph{48},
  291--300\relax
\mciteBstWouldAddEndPuncttrue
\mciteSetBstMidEndSepPunct{\mcitedefaultmidpunct}
{\mcitedefaultendpunct}{\mcitedefaultseppunct}\relax
\EndOfBibitem
\bibitem[McKemmish(2015)]{15Mc}
McKemmish,~L.~K. \emph{The Journal of Chemical Physics} \textbf{2015},
  \emph{142}, 134104\relax
\mciteBstWouldAddEndPuncttrue
\mciteSetBstMidEndSepPunct{\mcitedefaultmidpunct}
{\mcitedefaultendpunct}{\mcitedefaultseppunct}\relax
\EndOfBibitem
\bibitem[McKemmish and Gilbert(2015)McKemmish, and Gilbert]{15McGi}
McKemmish,~L.~K.; Gilbert,~A.~T. \emph{Journal of Chemical Theory and
  Computation} \textbf{2015}, \emph{11}, 3679--3683\relax
\mciteBstWouldAddEndPuncttrue
\mciteSetBstMidEndSepPunct{\mcitedefaultmidpunct}
{\mcitedefaultendpunct}{\mcitedefaultseppunct}\relax
\EndOfBibitem
\bibitem[Wolfram~Research(2018)]{wolfram}
Wolfram~Research,~I. {Mathematica}. 2018\relax
\mciteBstWouldAddEndPuncttrue
\mciteSetBstMidEndSepPunct{\mcitedefaultmidpunct}
{\mcitedefaultendpunct}{\mcitedefaultseppunct}\relax
\EndOfBibitem
\bibitem[Grev and Schaefer~III(1989)Grev, and Schaefer~III]{89GrSc}
Grev,~R.~S.; Schaefer~III,~H.~F. \emph{The Journal of chemical physics}
  \textbf{1989}, \emph{91}, 7305--7306\relax
\mciteBstWouldAddEndPuncttrue
\mciteSetBstMidEndSepPunct{\mcitedefaultmidpunct}
{\mcitedefaultendpunct}{\mcitedefaultseppunct}\relax
\EndOfBibitem
\end{mcitethebibliography}
\bibliographystyle{achemso}

\end{document}